\documentclass[10pt,journal,compsoc]{IEEEtran}

\usepackage[pdftex]{graphicx}
\usepackage{subcaption}
\usepackage{framed}
\usepackage{caption}
\DeclareGraphicsExtensions{.pdf,.jpeg,.png}

\usepackage{amsmath}
\usepackage{amssymb}
\usepackage{algorithmic}
\usepackage{tabularx}
\usepackage{array}
\usepackage{url}
\usepackage{listings}
\usepackage{multirow}
\usepackage{booktabs}
\usepackage{soul}

\usepackage{enumitem}

\newenvironment{conditions*}
  {\par\vspace{\abovedisplayskip}\noindent
   \tabularx{\columnwidth}{>{$}l<{$} @{\ : } >{\raggedright\arraybackslash}X}}
  {\endtabularx\par\vspace{\belowdisplayskip}}

\usepackage[autostyle=true]{csquotes}
\usepackage[english]{babel}
\usepackage[
 bibencoding=utf8
,backend=biber
,style=ieee
,minbibnames=1
,maxbibnames=2
,maxcitenames=1
,mincitenames=1
,eprint=false
,doi=true
,isbn=false
,url=true
]{biblatex}
\bibliography{verified_used_refs} 
\AtBeginBibliography{\footnotesize}

\DeclareSourcemap{
  \maps{
    \map{
      \step[fieldset=month, null]
      \step[fieldset=address, null]
      \step[fieldset=location, null]
      \step[fieldset=publisher, null]
      \step[fieldset=url, null]
      \step[fieldset=isbn, null]
    }
  }
}

\usepackage[
hidelinks,
  colorlinks=true,
  citecolor=blue,
  pdfstartview=Fit,
  breaklinks=true
]{hyperref}

\def\BibTeX{{\rm B\kern-.05em{\sc i\kern-.025em b}\kern-.08em
    T\kern-.1667em\lower.7ex\hbox{E}\kern-.125emX}}

\usepackage{tikz}

\usepackage{cleveref}
\crefname{lstlisting}{listing}{listings}
\Crefname{lstlisting}{Listing}{Listings}
\Crefname{figure}{Fig.}{Figs.}

\usepackage{flushend}

\begin{document}

\title{Tutorial on Flow-Based Network Traffic Classification Using Machine Learning}

\author{Adrián Pekár, Richard Plný, and Karel Hynek
\IEEEcompsocitemizethanks{\IEEEcompsocthanksitem A. Pekar is with the Budapest University of Technology and Economics, Hungary. A. Pekar is also with CUJO LLC, Hungary.
\IEEEcompsocthanksitem R. Plný and K. Hynek are with the Faculty of Information Technology, Czech Technical University in Prague.
\protect\\}
}

\IEEEtitleabstractindextext{%
\begin{abstract}
Modern networks carry increasingly diverse and encrypted traffic types that demand classification techniques beyond traditional port-based and payload-based methods. This tutorial provides a practical, end-to-end guide to building machine-learning-based network traffic flow classification systems. We cover the workflow from flow metering and dataset creation, through ground-truth labeling and feature engineering, to leakage-resistant experimental design, model training and evaluation, explainability, and deployment considerations. The tutorial focuses on supervised flow-based classification that remains effective under encryption and provides actionable guidance on algorithm selection, performance metrics, and realistic partitioning strategies, with emphasis on common real-world measurement artifacts and methodological pitfalls. A companion set of five Jupyter notebooks on GitHub implements the data-to-model workflow on real traffic captures, enabling readers to reproduce key steps. The intended audience includes researchers and practitioners with foundational networking knowledge who aim to design and deploy robust traffic classification systems in operational environments.
\end{abstract}

}

\maketitle

\IEEEdisplaynontitleabstractindextext

\IEEEpeerreviewmaketitle

\ifCLASSOPTIONcompsoc
\IEEEraisesectionheading{\section{Introduction}\label{sec:introduction}}
\else
\section{Introduction}
\label{sec:introduction}
\fi 

\IEEEPARstart{M}{odern} network management relies on granular visibility into traffic patterns to ensure security, performance, and reliability. In these environments, the fundamental unit of analysis is the \emph{network flow}---a sequence of packets sharing a common identifier, typically defined by the source and destination addresses and ports. These flows represent logical sessions between endpoints and can be unidirectional or bidirectional, serving as the basis for understanding how network resources are utilized. While flows may encompass unicast, multicast, or broadcast communications, their aggregation allows operators to move beyond individual packet inspection to a more holistic, session-oriented view of network activity.

To maintain operational health and situational awareness, operators must monitor capacity utilization and accurately classify the varied traffic types traversing their infrastructure. While endpoints could theoretically provide metadata about the traffic they generate, network operators rarely have administrative control over all connected devices, especially in the era of Bring Your Own Device (BYOD) and diverse IoT ecosystems. Consequently, operators must rely on independent methods to observe and analyze flows in transit. Traffic classification (TC) thus becomes a critical capability, enabling administrators to determine the class and purpose of network traffic even when the underlying devices or applications are unmanaged.

\subsection{Traffic Classification Use Cases}
\label{subsec:intro-tc-use-cases}

In practice, TC extracts meaningful information from network flows by mapping them to specific categories or application types. This process uncovers insights into application behavior, performance, and network status---attributes often hidden in raw packet streams. 
While traditional methods are challenged by modern protocols, TC increasingly leverages statistical and machine learning (ML) methods to remain effective against encryption. We group the essential use cases for traffic classification into five categories:

\subsubsection{Quality of Service Management}

Modern networks carry diverse traffic types---web browsing, video streaming, voice calls, file transfers, and more. Each type has unique requirements for bandwidth, delay sensitivity, and packet loss tolerance. For instance, video conferencing requires low latency, while file downloads need high throughput but can tolerate some delay.

Through accurate traffic classification, operators can identify these different traffic types and apply appropriate handling policies. This differentiation enables \emph{Quality of Service} (QoS) mechanisms that prioritize time-sensitive applications while ensuring fair resource allocation to lower-priority traffic. Traffic classification also supports \emph{traffic engineering}, where varying workloads are efficiently redistributed to maintain QoS and prevent congestion; flows with potential to overload links can be identified, re-routed, or throttled.

\subsubsection{Security and Threat Detection}

Network security increasingly relies on effective traffic classification. \emph{Denial of Service} (DoS) attacks attempt to disrupt services by overwhelming targets with traffic. These attacks come in two main forms: volumetric attacks that flood networks with excessive traffic, and application attacks that exploit software vulnerabilities to cause failures. Additionally, DoS attacks can be launched from multiple endpoints simultaneously, known as Distributed DoS (DDoS) attacks.

Traffic classification helps to detect both abnormal traffic volumes and unusual application usage patterns. This capability forms the foundation of network \emph{intrusion detection and prevention systems} (IDPS), which monitor network activity to identify potentially malicious behavior. These systems often employ anomaly detection, leveraging behavioral analysis of traffic patterns to uncover intrusions and other threats that signature-based methods cannot detect, such as zero-day exploits or attacks hidden within encrypted traffic.

\subsubsection{Network Planning and Business Operations}

Network operators must ensure efficient resource utilization while planning for future capacity needs. Traffic classification provides insights into usage patterns and growth trends across different application types, helping operators make informed investment decisions for \emph{capacity management}.

From a business perspective, TC also supports \emph{usage-based accounting} models where different traffic types may be billed at different rates (where legally permitted). It also helps verify compliance with \emph{terms of service} by identifying prohibited activities or excessive usage. Organizations can also use TC to identify unauthorized applications operating within enterprise networks, which present both security and compliance challenges.

\subsubsection{Troubleshooting and Performance Analysis}

When network performance issues arise, traffic classification significantly enhances \emph{troubleshooting} capabilities. Operators are required to investigate service-affecting incidents on their networks, which can be time-consuming. TC assists by providing access to current and historical data on what applications are traversing the network and what quality of service they are receiving. By identifying the specific applications affected and their normal behavioral patterns, operators can more quickly isolate and resolve problems.

\subsubsection{Regulatory and Legal Requirements}

Network operators often face legal obligations regarding data retention, lawful interception, and forensic analysis. Traffic classification helps operators meet these \emph{legal compliance} requirements by accurately identifying and documenting traffic types. Operators may be required to retain traffic measurement data for a set period in case it is needed for criminal investigations, and TC can provide the necessary classification of traffic types. This becomes particularly important in security incident investigations where understanding the nature of network communications may be legally mandated.

\subsection{Objectives and Scope}

Recent advancements in TC and ML have enabled the creation of increasingly sophisticated and powerful traffic classification systems. Yet, despite significant research advances, a substantial gap persists between theoretical developments and practical implementations in production networks~\cite{Azab2024}. This tutorial aims to bridge this gap by providing a practical and prescriptive guide to the complete pipeline shown in \Cref{fig:tc-ml-workflow}. We move beyond academic models to address the common challenges, measurement artifacts, and methodological pitfalls that practitioners regularly encounter.

Specifically, by the end of this tutorial, the reader should be able to:
\begin{itemize}
    \item design a flow-based measurement setup and anticipate common metering artifacts (\textit{e.g.}, sampling bias, timeouts, and network interface controller (NIC) offloading effects);
    \item obtain and assess ground-truth labels for flows using practical labeling methods (\textit{e.g.}, Deep Packet Inspection (DPI)-based and metadata-based labeling), and select strategies for handling limited or imperfect labels;
    \item construct leakage-resistant experimental protocols and dataset partitioning strategies suitable for traffic classification;
    \item train, evaluate, and interpret supervised ML classifiers for encrypted traffic using appropriate metrics and explainability methods;
    \item outline a deployment-ready classifier design with monitoring and drift considerations.
\end{itemize}

The tutorial focuses on supervised traffic classification problems such as application identification---a multi-class classification task distinguishing among potentially hundreds of application types based on their traffic signatures. As an increasing share of network communications becomes encrypted, we emphasize ML methods that remain effective under encryption, ensuring the presented techniques reflect modern network realities.

Our target audience includes researchers and engineers with foundational knowledge of networking concepts and basic familiarity with statistical methods. While some exposure to classification principles is beneficial, we introduce ML concepts as needed, making the material accessible to readers with varying levels of experience in this interdisciplinary domain. We select references to support key claims and point readers to useful entry points, rather than attempting comprehensive literature coverage.

We focus on principles that form the backbone of ML-based traffic classification while selectively addressing state-of-the-art techniques suited to contemporary practice. This measured approach emphasizes core concepts without sacrificing exposure to sophisticated methods---all within a manageable scope that prioritizes clarity and utility over encyclopedic depth, supporting both academic study and real-world deployment.

\subsection{Paper Structure}
\label{subsec:intro-paper-structure}

The remainder of this tutorial is organized as follows. \Cref{sec:tc-techniques} provides a brief overview of the evolution of traffic classification techniques, establishing why ML-based approaches have become dominant. The core of this work then covers the three main stages of the ML pipeline: data collection (\Cref{sec:data-collection}), data preparation (\Cref{sec:data-preparation}), and ML model development (\Cref{sec:machine-learning}). Moving beyond theory, this tutorial also includes a practical component: a series of five Jupyter notebooks that implement the complete data-to-model pipeline (data collection, preparation, modeling, and evaluation) using real network traffic captures. Model deployment considerations are discussed in \Cref{sec:machine-learning}, while the notebooks focus on the data-to-model workflow. These hands-on exercises, described in \Cref{sec:practical-part}, allow readers to apply the concepts presented in the preceding sections and experience common pitfalls firsthand. \Cref{sec:conclusion} concludes the tutorial.

\section{The Evolution of Traffic Classification Techniques}
\label{sec:tc-techniques}

TC techniques have evolved significantly over the years in response to shifts in network protocols and usage patterns, prompting the development of both new methods and refinements of the existing ones. In this section, we provide a brief overview of the TC research evolution and outline the current landscape of TC approaches, which vary widely depending on the specific use case.

\subsection{Port and IP Address-Based Classification: The Naive Approach}

In the early days of networking, traffic classification was relatively straightforward. The Internet landscape was less complex, and applications generally adhered to well-known transport-layer port numbers registered with the Internet Assigned Numbers Authority (IANA). Network operators could identify services through static port-to-application mappings and, when needed, maintain IP-to-service mappings associating known server ranges with specific providers.

These simple list-based classification methods enabled early implementations of Quality of Service (QoS) and basic traffic management. However, both approaches have become largely obsolete as the network ecosystem evolved. The widespread adoption of HTTP/HTTPS as a universal transport mechanism for diverse application types---often multiplexed within a single connection---has undermined the reliability of port-based identification. Similarly, the shift away from application-specific IANA-registered ports toward dynamic port allocation, coupled with the prevalence of shared cloud infrastructures where a single IP address may host multiple unrelated services, has rendered such naive classification methods ineffective in modern networks~\cite{4738466,6786614}.

\subsection{Deep Packet Inspection}
\label{subsec:tc-dpi}

As port-based methods declined in effectiveness, \textit{Deep Packet Inspection} (DPI) emerged as a widely adopted approach~\cite{6644335}. DPI analyzes the payloads of network packets to perform a range of tasks, from basic application-protocol identification to detecting specific attacks and instances of service misuse. However, DPI also raises privacy concerns, as it can be used for surveillance and censorship~\cite{Dyer:2013:PMM:2508859.2516657,Wang:2015:STN:2810103.2813715,7782699}.

One DPI method is pattern matching. \emph{Patterns} that uniquely identify a protocol are derived from the protocol specification. Traffic is classified by matching the payload against known patterns and identifying the corresponding protocol. Patterns can be exact strings, byte sequences, or regular expressions. For example, OpenVPN can be detected through its 5-bit opcode message types~\cite{VPNOpCodeDetection}. Another DPI method uses protocol parsers and finite state machines, also called state-aware decoders, which can detect out-of-order packets, malformed packets, and possible protocol-targeted exploits.

In recent years, DPI has been facing major limitations due to the widespread adoption of traffic encryption and the resulting reduction in on-path visibility~\cite{rfc8546}. While DPI can still leverage unencrypted metadata or be applied in environments where traffic is decrypted (\textit{e.g.}, enterprise TLS inspection proxies), it is often ineffective for passive on-path observers when decryption is unavailable. Consequently, the importance of flow-based traffic classification has grown significantly~\cite{alwhbi2024encrypted}.

\subsection{Machine Learning}
\label{subsec:tc-traffic-classification}

Machine learning started gaining adoption in networking in the early 2000s~\cite{nguyen2009survey}. Today, ML-based approaches are widely used and have become central to traffic classification research and practice, driven in part by their ability to model statistical patterns that remain observable under encryption~\cite{alwhbi2024encrypted,luxemburk2023encrypted}. Initially, \textit{shallow models} such as Decision Trees, $k$-Nearest Neighbors, Naive Bayes, Support Vector Machines, and later Random Forests were predominantly used, with research efforts focusing on feature selection and engineering. Early approaches relied on packet and payload-based features. However, as data encryption became standard practice, payload-based features lost their relevance. This shift led to the increased use of flow-based features and features derived from packet headers. Despite extensive research, no standardized feature set has emerged in the TC domain, and a wide range of feature vectors is being actively explored~\cite{nguyen2009survey}. Furthermore, some studies, such as~\cite{arp2022and}, have identified common pitfalls associated with applying ML to TC tasks. 

In the mid-2010s, \textit{deep learning} approaches became increasingly common in traffic classification research, including convolutional neural networks operating directly on packet sequences~\cite{Wang2017,alwhbi2024encrypted}. In addition, advanced methods such as metric learning, meta-learning, and other sophisticated techniques have also been investigated. Despite the lack of a standardized feature set, information extracted from the first $N$ packets of a flow---such as packet sizes, inter-packet arrival times, and directions---has become increasingly prevalent in recent approaches.

\begin{figure}[t]
	\centering
	\includegraphics[width=\columnwidth]{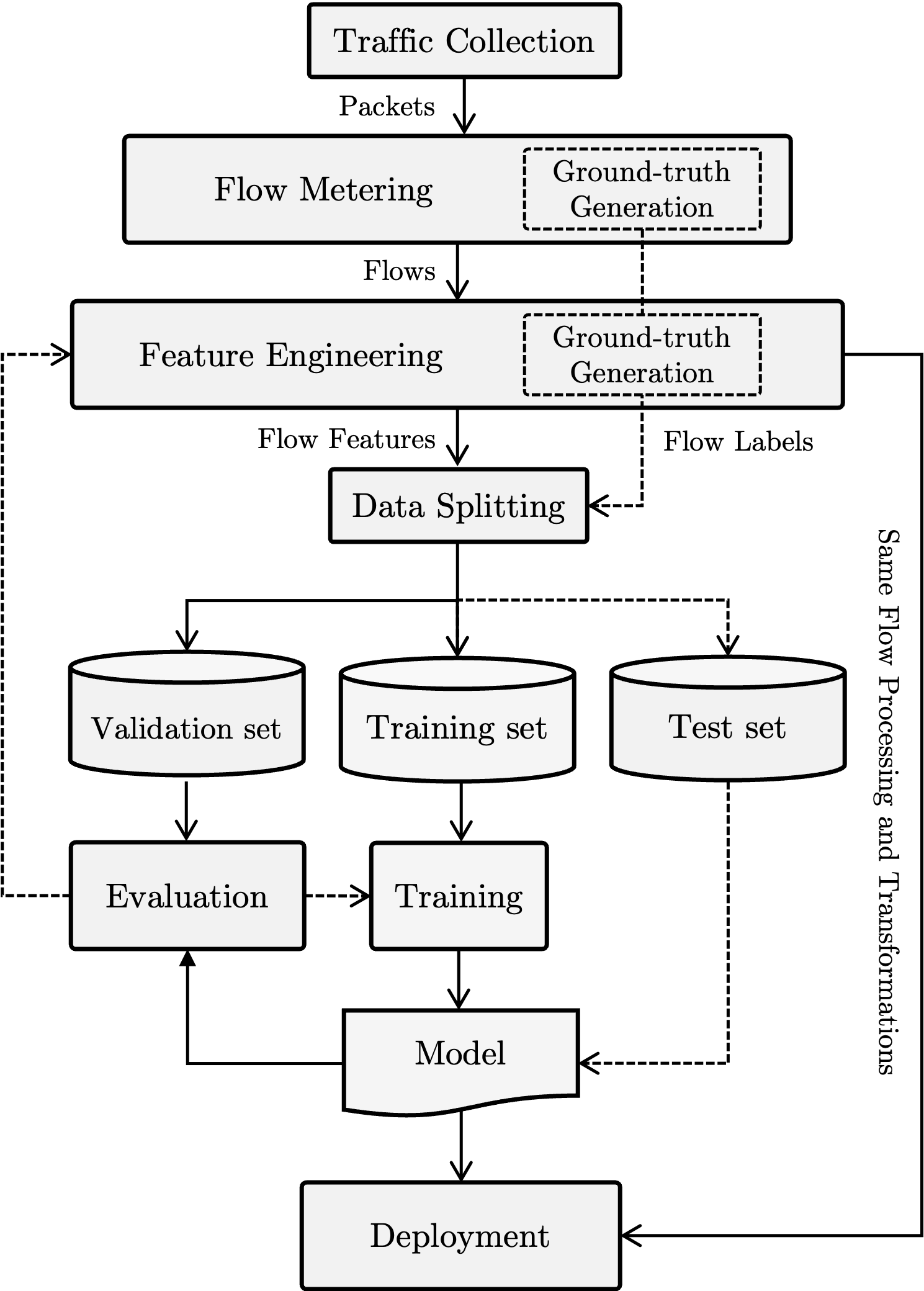}
	\caption{Overview of the ML workflow for traffic classification}
	\label{fig:tc-ml-workflow}
\end{figure}

Today, ML is applied in various networking approaches such as traffic fingerprinting~\cite{sibyencrypted,hayes2016k}, traffic correlation~\cite{Li2021,Nasr2018,Zhu2024}, and anomaly detection~\cite{goswami2023unsupervisedmodelselectiontimeseries,tadgan,crae}. Nevertheless, to maintain the focus and conciseness of this tutorial, we concentrate primarily on supervised traffic classification---assigning category labels to traffic flows based on learned patterns.

Designing and deploying an ML classifier is a non-trivial process. As illustrated in \Cref{fig:tc-ml-workflow}, the process begins with obtaining a high-quality dataset, followed by rigorous data preparation and the design of an evaluation protocol (data splitting). Subsequent steps, including model training, hyperparameter tuning, and final model selection, are time-intensive but heavily influence both the predictive performance of the model and its operational deployability. The following sections provide the necessary theoretical foundation alongside practical guidance on how to overcome these challenges.

\section{Data Collection}
\label{sec:data-collection}

Network traffic measurement is the foundation of any classification system, balancing a core trade-off between \textit{fidelity}---the accuracy and granularity of captured data---and \textit{scalability} in high-throughput environments. Detailed measurements offer richer information but at high computational and storage costs, while coarse measurements risk missing patterns essential for accurate classification.

This tutorial focuses on \textit{traffic flow measurement}, where packets are captured and aggregated into flow records enriched with statistical features for ML. We outline how raw packets are transformed into analysis-ready flow records, emphasizing key stages, technical considerations, and trade-offs relevant to both research and operational practice.

\subsection{From Packets to Flows}

\begin{figure}[t!]
	\centering
	\includegraphics[scale=0.72]{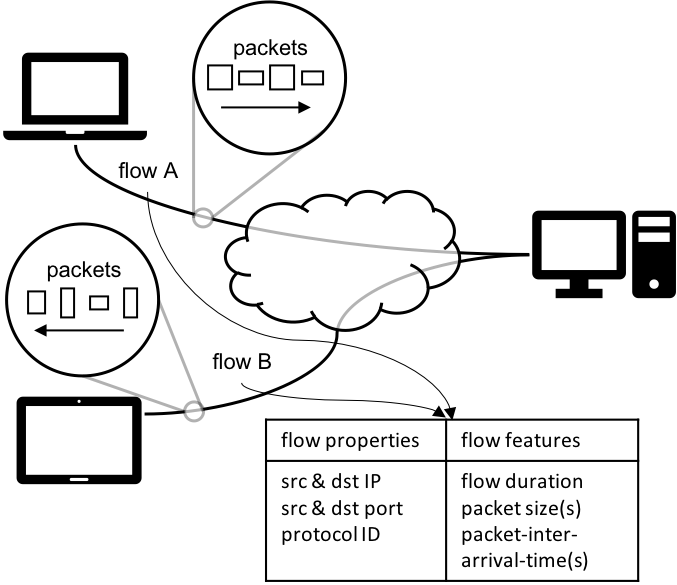}
	\caption{The relationship between packets, flows, flow properties, and features. Flow properties (the 5-tuple) define which packets belong to a flow, while flow features are statistical measures calculated from those packets, forming the input for classification algorithms.}
	\label{fig:flow}
\end{figure}

In general, networks transmit data as individual packets. While packet-level analysis is sometimes needed for detailed forensic work, it is often inefficient for higher-level behavioral analysis. Usually, ML-based traffic classification instead relies on the abstraction of a \textit{flow}. A flow groups packets into structured records that capture the \textit{conversations} between endpoints. As shown in \Cref{fig:flow}, this abstraction forms the basis of modern traffic analysis. Grouping packets into higher-level records offers several key advantages:

\begin{LaTeXdescription}
    \item[Data Reduction:] Aggregating packets into flows can typically reduce data volume by orders of magnitude, making analysis computationally feasible~\cite{Hofstede2014}.
    \item[Behavioral Feature Extraction:] It enables the calculation of statistical features (\textit{e.g.}, timing regularities, size distributions) that emerge from the relationship between multiple packets and are invisible at the single-packet level.
    \item[Encryption Resilience:] Flows are defined using network-layer (L3) and, when observable, transport-layer (L4) headers and metadata. While encryption, encapsulation, and tunneling can reduce the information visible to an on-path observer, flow records often retain sufficient statistical structure for traffic classification~\cite{rfc8546,alwhbi2024encrypted}.
\end{LaTeXdescription}

\subsubsection*{Defining a Flow}
A flow is a set of IP packets passing an observation point in the network during a certain time interval~\cite{rfc7011}. All packets belonging to a flow have a set of common properties. Although such a definition allows arbitrary aggregation of packets by the common properties, the standard aggregation key (or flow key) is the following 5-tuple:

\begin{itemize}
    \item Source IP address,
    \item Destination IP address,
    \item Source port,
    \item Destination port, and
    \item Transport Protocol (\textit{e.g.}, TCP, UDP).
\end{itemize}

All packets matching this 5-tuple within a certain time window are considered part of the same flow. From this group of packets, \textit{flow features} are calculated. The flow features, which usually contain statistical information, are then used for further analysis. Nevertheless, flows can also be enriched with selectively extracted application-layer metadata that may be observable at capture time (\textit{e.g.}, TLS handshake fields such as the SNI, when present). Depending on protocol versions and privacy extensions (\textit{e.g.}, TLS~1.3 and Encrypted ClientHello (ECH)), such fields may be unavailable, and many operational settings treat payload as inaccessible.

\subsection{The Flow Generation Pipeline}
\begin{figure}[ht]
	\centering
	\includegraphics[scale=0.63]{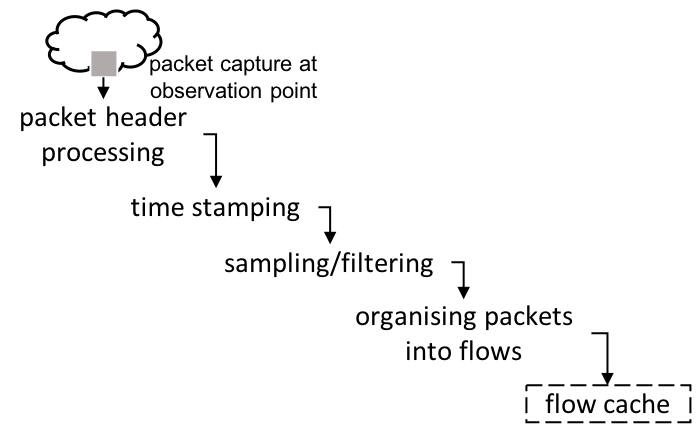}
	\caption{The flow generation pipeline. The process involves four key stages: (1) packet capture, (2) precise timestamping, (3) optional packet selection, and (4) aggregation into flow records stored in a flow cache.}
	\label{fig:packet-to-flow}
\end{figure}

The process of transforming raw network traffic into structured flow records follows a well-defined pipeline, as illustrated in \Cref{fig:packet-to-flow}. This process, often referred to as \textit{flow metering}, involves several key stages.

\paragraph*{\textbf{Packet Capture}} Traffic is first intercepted at an observation point. The technology used for capture has significant implications for both fidelity and cost. Options range from high-throughput dedicated hardware probes to flexible software libraries (\textit{e.g.}, libpcap), modern programmable data planes (\textit{e.g.}, P4, DPDK), and embedded export protocols like NetFlow/IPFIX found in commercial routers.

\paragraph*{\textbf{Timestamping}}
Each captured packet is marked with a precise timestamp. This is a critical step, as the quality of all time-based features (\textit{e.g.}, flow duration, inter-arrival times) depends on it. High-speed networks demand microsecond-level precision. Consider packets arriving 100 microseconds apart: if the timestamping resolution is only 1 millisecond, these packets appear simultaneous, collapsing their inter-arrival times to zero and distorting timing-based features. Hardware-assisted timestamping generally provides this precision more reliably than software-based methods, which are subject to OS scheduling delays and interrupt latency.

\paragraph*{\textbf{Optional Packet Selection}}
In high-throughput environments, processing every single packet can be prohibitively expensive. In such cases, a selective approach is often employed to make the task tractable. This can involve statistical \textit{sampling} (\textit{e.g.}, selecting 1-in-N packets) or deterministic \textit{filtering} (\textit{e.g.}, capturing only TCP traffic on port 443). However, this step must be used with extreme caution, as it introduces significant measurement bias. For instance, 1-in-N sampling will artificially inflate measured Packet Inter-Arrival Times (PIAT) and can cause very short flows to be missed entirely. These biases must be understood and potentially corrected in subsequent analysis.

\paragraph*{\textbf{Flow Record Creation}}

Finally, the (potentially sampled) packets are aggregated into flow records in a \textit{flow cache}. An active flow record is created for each new 5-tuple observed. As subsequent packets matching that 5-tuple arrive, the record's statistical features (\textit{e.g.}, packet count, total bytes, timestamp of last seen packet) are updated. A flow record is considered complete and is exported for analysis when it meets a certain condition, such as observing a connection teardown (\textit{e.g.}, TCP FIN (finish) or RST (reset) flags) or being idle (or inactive) for some time (\textit{e.g.}, no packets seen for 30 seconds). The resulting collection of exported flow records forms the dataset for our classification task.

\subsection{Flow Directionality}

\begin{figure}[ht]
	\centering
	\includegraphics[scale=0.70]{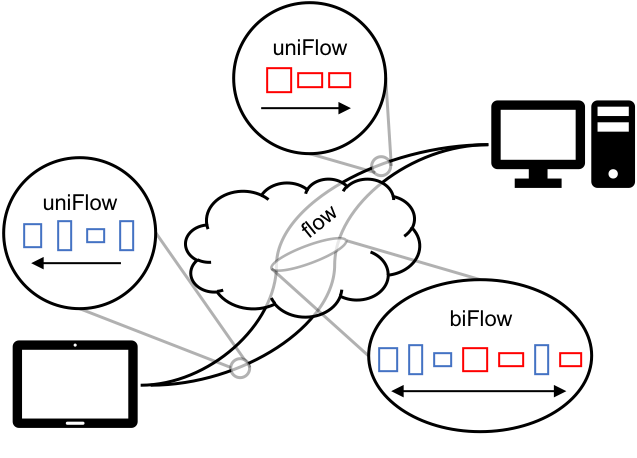}
	\caption{Unidirectional versus bidirectional flow modeling. Unidirectional modeling creates separate records for each direction. Bidirectional modeling captures the entire two-way exchange in a single, comprehensive record, which is commonly preferred for modern ML-based classification.}
	\label{fig:u-b-flow}
\end{figure}

A crucial design decision in any flow metering process is how to represent the inherently bidirectional nature of network communication. A logical session between a client and a server involves traffic in both the forward (client-to-server) and backward (server-to-client) directions. The way these directions are captured in the final flow records fundamentally shapes what patterns can be analyzed. As illustrated in \Cref{fig:u-b-flow}, this leads to two primary modeling approaches.

\subsubsection*{Unidirectional Flows (UniFlows)}

The most basic concept is the unidirectional flow. Here, a flow record is built from packets traveling in \textit{one direction} only. To capture a conversation between endpoints A and B, two separate UniFlow records are needed: one for A$\rightarrow$B and another for B$\rightarrow$A. This representation is the most common, since hardware switches and routers typically export only unidirectional flows, and it achieves this with minimal performance cost. Its main drawback, however, is that the relationship between the two directions is lost at the record level. To reconstruct the full conversation, a practitioner must perform additional post-processing to pair the corresponding unidirectional records, a task that is both complex and error-prone.

\subsubsection*{Bidirectional Flows (BiFlows)}

This second concept views the entire two-way exchange between endpoints A and B as a single, indivisible unit. Here, features are calculated by \textit{aggregating packets from both directions} (from A$\rightarrow$B and B$\rightarrow$A) into one record. This approach creates a larger record that simultaneously contains three distinct sets of features:
\begin{itemize}
    \item A set of features representing the \textit{forward UniFlow} perspective (\textit{e.g.}, prefixed `fwd\_` or `src2dst\_`), calculated only from A$\rightarrow$B packets.
    \item A set of features representing the \textit{backward UniFlow} perspective (\textit{e.g.}, prefixed `bwd\_` or `dst2src\_`), calculated only from B$\rightarrow$A packets.
    \item A set of features representing the \textit{aggregated BiFlow} perspective (\textit{e.g.}, prefixed `bidirectional\_` or using non-prefixed names like `flow\_duration`), calculated from all A$\leftrightarrow$B packets.
\end{itemize}

The main motivation for creating BiFlows is to combine both directions into a single record, reducing bandwidth usage and computational cost during analysis. In practice, however, BiFlows are not always feasible. Environments with asymmetric routing or unidirectional network taps may expose only one traffic direction, forcing a fallback to the UniFlow model. As a result, some flow records contain only forward-direction fields. To recover BiFlows, the monitoring pipeline must later perform flow-stitching, an additional post-processing step.

\subsection{Flow Cache System}

In an operational environment, a flow metering system must operate at line rate, processing millions of packets per second. At the heart of this high-performance task is the \textit{flow cache}---a highly optimized, in-memory data structure that stores the state of all active flow records. The design of this cache is a critical engineering challenge that directly impacts both the quality of the final features and the overall throughput of the measurement system.

\begin{figure}[h!]
\begin{center}
    \includegraphics[scale=0.63]{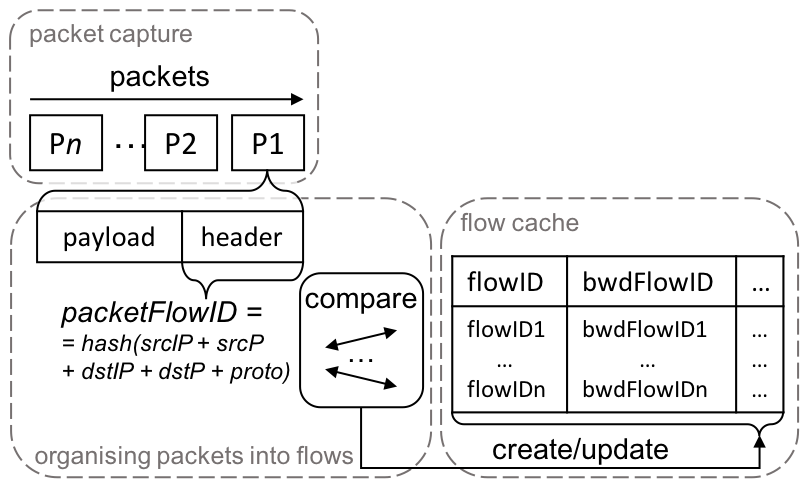}
    \caption{The flow cache lookup process. A packet's 5-tuple is hashed to create a flowID, which is used as a key to search the flow cache. If a match is found, the existing flow record is updated; otherwise, a new record is created.}
\label{fig:fwd}
\end{center}
\end{figure}

As illustrated in \Cref{fig:fwd}, the process begins when a network packet arrives. The system must rapidly determine if this packet belongs to an existing flow in the cache or if it represents the start of a new one. This is achieved through a high-speed, hash-based lookup mechanism.

\subsubsection*{The Packet Management Process}

The core workflow for each incoming packet involves three steps:
\begin{enumerate}
    \item \textit{Flow Key Extraction:} The system first extracts the 5-tuple properties (source/destination IPs and ports, protocol) from the packet's header.
    \item \textit{Hash Computation:} The extracted Flow Key is then fed into a specialized hash function (\textit{e.g.}, Murmur3, CRC32) to compute a single hash value, the \emph{flowID}. This flowID serves as the lookup key into the hash table (flow cache). The use of a hash table is crucial, as it provides an average-case time complexity of $O(1)$ for lookups, making it scalable to millions of concurrent flows.
    \item \textit{Cache Action (Update or Create):} The flowID is used to query the cache. If a matching entry is found, the packet is assigned to that flow and its features are updated (\textit{e.g.}, packet counters, last-seen timestamp). If no match exists, a new flow record is created and initialized from the packet. If the hash table slot is already occupied by a different flowID, a collision handling routine is triggered. Commonly, the older flow is evicted in favor of the new one, though many alternative strategies and hash table designs exist.
\end{enumerate}

\subsubsection*{Bidirectional Packet Management Process}

The simple lookup process becomes more complex when generating the BiFlow records discussed previously. The system must be able to recognize that a packet traveling from B$\rightarrow$A belongs to the same session as a packet from A$\rightarrow$B. Two primary techniques are used to solve this association problem at line rate.

\begin{itemize}
    \item \textit{Dual Hash Lookup:} This method computes two hashes per packet: a \emph{forward flowID} from the original 5-tuple, and a \emph{reverse flowID} from the 5-tuple with source and destination pairs swapped. The system then checks both flowIDs in the flow cache. A hit on either identifies the packet as part of an existing bidirectional flow, with the matching hash revealing its direction. This enables updating the appropriate directional features (`fwd\_` or `bwd\_`). The trade-off is the added cost of two hash computations and two cache lookups per packet.

    \item \textit{Canonical Key Lookup:} This approach forces packets from both directions of a flow to hash to the same value, requiring only one lookup. It does so by creating a \textit{canonical representation} of the 5-tuple, where fields such as IP addresses are numerically ordered (\textit{e.g.}, placing the lower IP and port first). This ensures A$\rightarrow$B and B$\rightarrow$A packets share the same flowID. While it eliminates an extra lookup, it adds overhead from normalizing the key for each packet and requires post-lookup checks to determine the packet’s direction before updating the `fwd\_` or `bwd\_` features.
\end{itemize}

The choice between these two methods involves a trade-off between the cost of multiple cache lookups versus the cost of key normalization and post-lookup comparison. The optimal solution depends on the specific performance characteristics of the target hardware and software environment.

\subsection{Flow Lifecycle}
A flow record is not static and follows a predetermined lifecycle. When a packet with a new flow key arrives, a flow record is created; when a packet with an existing key arrives, the corresponding record is updated. The most difficult question is when a flow should be considered \textit{complete} and exported for analysis. This process is called \textit{flow expiration}.

Modern flow meters implement several complementary expiration mechanisms. The most common are based on timeouts, which ensure that no flow record remains in the cache indefinitely.

\subsubsection*{Idle Timeout}
This is the most common expiration trigger. A flow record is terminated if no new packets matching its 5-tuple are observed for a specified time period. This mechanism is essential for handling flows that end without an explicit protocol signal (\textit{e.g.}, natural end of UDP flows).
\begin{itemize}
    \item \textit{Typical Values:} 15-30 seconds for high-throughput networks, or 60-120 seconds for lower-volume environments~\cite{Hofstede2014,rfc3954}.
    \item \textit{Classification Impact:} This parameter creates a direct trade-off. Shorter timeouts provide more timely data but risk fragmenting application sessions that have natural pauses in their traffic, potentially disrupting behavioral features. Longer timeouts better capture complete sessions but increase the risk of incorrectly merging two separate, consecutive communications that reuse the same 5-tuple.
\end{itemize}

\subsubsection*{Active Timeout}
This mechanism ensures that even very long-lived, continuously active flows are eventually exported for analysis. A flow is forcibly terminated after it has been active for a maximum duration, regardless of ongoing packet arrivals. This is critical for segmenting persistent connections (\textit{e.g.}, streaming sessions, VPN tunnels) into manageable, analyzable chunks.
\begin{itemize}
    \item \textit{Typical Values:} 5-30 minutes, depending on memory constraints and the need for timely analysis~\cite{Hofstede2014,rfc3954}.
    \item \textit{Classification Impact:} This directly affects the observation window for long-term behavioral patterns. Active timeouts that are too short might significantly degrade the classification accuracy for streaming applications by preventing the model from observing the full, characteristic timing patterns.
\end{itemize}

\begin{figure}[!t]
    \centering
    \includegraphics[width=0.8\columnwidth]{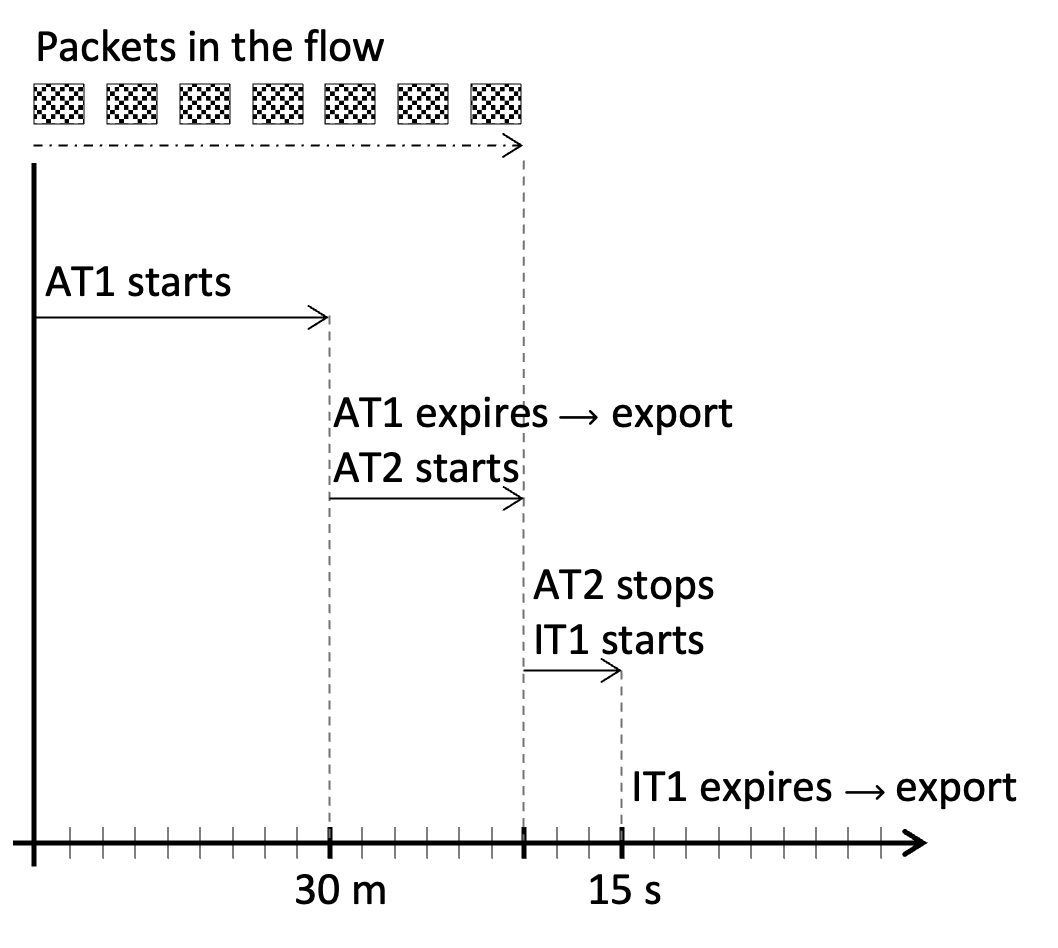}
    \caption{The interplay of active (AT) and idle (IT) timeouts. A continuous packet stream is first segmented by the expiration of AT1. When the stream stops, the final record is exported upon the expiration of IT1.}
\label{fig:timeout-example}
\end{figure}

The interplay between these two timeout mechanisms is illustrated in \Cref{fig:timeout-example}. A single, long-running stream of packets is first segmented by the active timeout (AT1), which triggers an export and the start of a new record. When the packet stream eventually stops, the idle timeout (IT1) begins and later triggers the final export.

\subsubsection*{Protocol-Based Expiration}
Beyond timeouts, flow meters can leverage explicit protocol signals to identify natural session boundaries. For TCP, observing a packet with the FIN or RST flag is a strong indicator that a connection is terminating, allowing for immediate and accurate flow expiration. This method is highly effective as it preserves the true lifecycle of the application's session, leading to higher-quality features.

\subsubsection*{Resource-Constrained Expiration}
Finally, flow meters must incorporate a fail-safe mechanism to ensure stability under extreme load. When the flow cache is near its memory limit due to excessive concurrent flows, a resource-constrained expiration policy is applied. This mechanism frees memory by expiring the oldest or least-recently seen flows, allowing new entries to be admitted. While critical for preventing system failure, such forced expirations can introduce measurement artifacts during high-traffic periods, as flows may be prematurely terminated, degrading the quality of their statistical features.

\subsubsection*{The Impact of Flow Timeouts}

\begin{figure}[t!]
	\centering
	\includegraphics[width=\columnwidth, trim=0 3cm 0 0, clip]{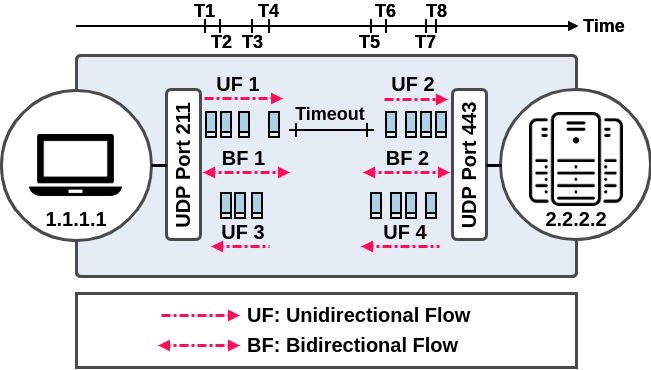} 
	\caption{Illustration of flow splitting due to an idle timeout. The flow meter creates the first bidirectional flow record (`BF 1`), which is composed of the forward unidirectional component (`UF 1`) and the backward unidirectional component (`UF 3`). After a period of activity, the communication pauses. When this pause exceeds the idle timeout, the meter finalizes and exports the record. When the application resumes, the meter treats this as a new session and creates a second bidirectional record (\textit{BF 2}), composed of its own unidirectional components (`UF 2` and `UF 4`).}
	\label{fig:flow-timeout-split}
\end{figure}

The timeout mechanisms discussed previously have a profound and practical implication: a single, long-lived application session can be \textit{fragmented} into multiple, sequential flow records in the dataset. This phenomenon (illustrated in \Cref{fig:flow-timeout-split} and contextualized in \Cref{tab:flow-record}) is not an error but a fundamental consequence of how flow meters manage resources. A single communication is split into two subflows, `Flow 1` and `Flow 2`, where the `fwd\_` features of the first record are computed from the packets in `UF 1`, while the `fwd\_` features of the second record are computed from the packets in `UF 2`. The backward (`bwd\_`) and bidirectional (`total\_`) features are derived analogously. As a result, the subsequent analysis needs to deal with two distinct flow records for what was a single logical conversation.

This flow-splitting behavior poses a significant challenge. The dataset now consists of multiple partial flow records, each capturing different properties. For example, the first record may miss the connection-termination signature, while the second record lacks the initial connection-setup phase, such as the TCP handshake. A subsequent analysis needs to then account for these measurement artifacts.

\subsection{Flow Feature Computation and Selection}
The essence of flow-based analysis lies in aggregating raw packet streams into statistical summaries and features. Classical flow records typically provide basic information, such as the number of transmitted bytes and packets in both forward and backward directions. However, flows can be enriched with additional features tailored to specific use cases, including ML-based analysis. These flow features targeting ML are commonly organized into four primary categories, each capturing a distinct aspect of communication behavior.

\paragraph*{\textbf{Packet Size Features}} They capture the distribution of packet lengths within a flow. These features are highly discriminative, as different applications exhibit distinct data transfer patterns. For instance, bulk transfers and video streaming typically use large packets near the Maximum Transmission Unit (MTU) to maximize throughput, whereas interactive applications such as SSH or remote desktop rely on smaller packets that directly reflect user activity. Beyond their discriminative power, packet size features are also valued for their stability: unlike timing-based metrics, they are less influenced by transient conditions such as congestion. Although phenomena like fragmentation or tunneling encapsulation can still modify packet sizes mid-path, this relative robustness makes size-based features among the most important in modern traffic classification approaches~\cite{luxemburk2023encrypted}.

\begin{table}[t!]
  \centering
  \caption{Example of features for two bidirectional flow records. Each record maintains separate statistics for the forward and reverse directions (representing the UniFlow perspectives), alongside combined totals (representing the BiFlow perspective), enabling classifiers to learn from the crucial relationship between the directional components.}
  \label{tab:flow-record}
  \begin{tabular}{rll}
    \toprule
    \textbf{Flow prop. and features} & \textbf{BiFlow 1} & \textbf{BiFlow 2}\\
    \midrule
    Id & 1& 2\\
    Source IP Address & 1.1.1.1& 2.2.2.2\\
    Destination IP Address & 2.2.2.2& 1.1.1.1\\
    Source Port & 211& 443\\
    Destination Port & 443& 211\\
    Transport Protocol Id & 17& 17\\
    \textit{fwd\_packet\_count} & 4& 4\\
    \textit{bwd\_packet\_count} & 3& 4\\
    \textit{total\_packet\_count} & 7& 8\\
    Flow Start Time & T1& T5\\
    Flow End Time & T4& T8\\
    \textit{fwd\_duration} & (T4 - T1)& (T7 - T5)\\
    \textit{bwd\_duration} & (T3 - T2)& (T8 - T6)\\
    \textit{flow\_duration} & (T4 - T1)& (T8 - T5)\\
  \bottomrule
\end{tabular}
\end{table}

\paragraph*{\textbf{Timing Features}} They capture the temporal rhythm of communication, with PIAT statistics being especially informative. Different applications produce distinct timing signatures: streaming media often yields regular, near-constant PIATs dictated by codec bitrates, whereas interactive sessions generate irregular, bursty patterns reflecting human activity. The strength of timing features, however, is also their main challenge---they are highly sensitive to network conditions. Unlike the relative stability of packet sizes, observed timing reflects both the application’s intrinsic signature and noise from network effects such as queuing delay, jitter, congestion or simply the position of the flow meter in the network.

\paragraph*{\textbf{Volume Features}} They aggregate metrics that quantify the overall size and scale of a flow. Simple features like total packet and byte counts, flow duration, and the ratio between forward and backward volumes are computationally inexpensive yet surprisingly effective. Derived metrics like bytes-per-packet (average packet size) and packets-per-second are particularly useful as they provide normalized measures of a flow's characteristics.

\paragraph*{\textbf{Protocol Context Features}} They leverage available protocol header information. The protocol identifier (TCP vs. UDP) provides a strong clue about the application's reliability requirements. TCP flags (when observable) reveal connection behavior patterns. Similarly, while port numbers are no longer a reliable primary identifier, they can still provide useful context when combined with the behavioral features described above.

\subsection{Flow Labeling and Obtaining Ground Truth}
\label{sec:ground-truth}

Supervised ML critically depends on the quality of its training data, which in the context of traffic classification requires flow records annotated with accurate ground-truth labels. Obtaining such reliable labels is usually the biggest challenge in developing practical classification systems, because network traffic lacks inherent application identifiers---otherwise, the very task of traffic classification would be unnecessary. This section examines the methodologies used to obtain these labels and discusses the trade-offs of each approach.

\begin{table*}[!ht]
\scriptsize
\centering
\caption{Comparison of Primary Ground-Truth Labeling Methods for Traffic Classification}
\label{tab:labeling_comparison}
\begin{tabular}{@{}p{2.5cm} p{3cm} p{2.5cm} p{2.5cm} p{2cm} p{3cm}@{}}
\toprule
\textbf{Method} & \textbf{Typical Accuracy} & \textbf{Scalability / Throughput} & \textbf{Deployment Difficulty} & \textbf{Privacy Concerns} & \textbf{Realism / Diversity} \\
\midrule
Deep Packet Inspection (DPI) & High for known, unencrypted; Low for encrypted/novel & Low (computationally expensive) & Moderate (requires traffic access point) & Moderate (sees payload snippets) & High (if applied to real traffic) \\
\midrule
Controlled Environment & Perfect (within lab context) & N/A (generates data, does not process live) & High (setting up diverse apps/conditions) & Low (controlled data) & Low (struggles to capture real-world complexity/mix) \\
\midrule
Port-Based & Very Low; Moderate for specific legacy protocols & Very High (trivial computation) & Very Low (uses existing flow data) & Low (uses header info only) & High (if applied to real traffic) \\
\midrule
Active Labeling & Moderate (depends on target responsiveness and data freshness) & Low (active probes and lookups are computationally heavy) & High (requires active querying or integration with external platforms) & High (involves scanning or probing remote hosts) & Moderate (data may be outdated in public databases) \\
\midrule
External Intelligence Providers & Moderate to High (depends on data quality and source reliability) & Low to Moderate (API can be rate-limited, bulk syncs scale well) & Moderate (requires API access and periodic synchronization) & Low (no direct interaction with remote hosts) & Moderate (data may be outdated or limited by coverage) \\

\bottomrule
\end{tabular}
\end{table*}

\subsubsection{Primary Labeling Methodologies}
Several approaches have been developed to generate ground truth, each occupying a different point on the spectrum of accuracy, scalability, and operational feasibility. \Cref{tab:labeling_comparison} summarizes these trade-offs. The methods can also be combined and cross‑validated to maintain labeling consistency, while voting schemes or confidence scoring based on the chosen method and traffic characteristics can further enhance label reliability.

\paragraph*{\textbf{Deep Packet Inspection (DPI)}}
DPI remains a widely used approach for generating labels by analyzing packet payloads for application-specific signatures. However, its effectiveness has significantly declined with the growing prevalence of encryption and the emergence of new applications. In addition, the high computational cost of DPI limits its scalability, restricting its use to smaller data subsets for preliminary labeling. DPI also encompasses the analysis of unencrypted parts of communication within protocols typically associated with encryption. For instance, the TLS Server Name Indication (SNI) extension can reveal the requested hostname when transmitted in plaintext; however, SNI may be absent or encrypted by privacy extensions (\textit{e.g.}, Encrypted ClientHello (ECH)), and therefore cannot be assumed to be consistently observable.

\paragraph*{\textbf{Controlled Environment Testing}}
An alternative is to generate traffic in a controlled laboratory environment where the source application is definitively known. This method produces perfect, high-confidence labels and allows for reproducible experiments. The significant drawback, however, is the transferability and \textit{realism gap}~\cite{liu2024operationalizing}. Lab-generated traffic often fails to capture the immense diversity and complexity of real-world network conditions and user behaviors. Models trained exclusively on such data often experience a significant drop in performance when deployed in production.

\paragraph*{\textbf{Port-Based Labeling}}
While obsolete as a primary classification technique, using port numbers can still serve as a supplementary labeling method for a small set of legacy, well-behaved protocols (\textit{e.g.}, SSH on port 22, SMTP on port 25). It is computationally trivial and can provide a useful, low-confidence baseline for these specific services, but it is entirely unreliable for the vast majority of modern applications that use dynamic ports or tunnels through standard web ports.

\paragraph*{\textbf{Active Labeling}}
When Deep Packet Inspection is not possible due to encryption, an alternative approach is to actively query the communication target. In cases of unknown or encrypted traffic, sending a similar connection request to the destination server can help reveal useful metadata such as domain names, certificates, or the type of service hosted on the server. However, this method has significant drawbacks. It consumes substantial computational resources, and it raises ethical concerns because it involves active Internet host scanning. A less intrusive alternative is to rely on publicly available scanning databases such as Shodan or Censys. These platforms perform continuous large-scale Internet scanning and provide detailed information about hosted services, open ports, and certificates. The main limitation of this approach is data obsolescence: information in Shodan or Censys can be several days old (Shodan states that it scans each host at least once a week), which may lead to inaccurate labels when host configurations change rapidly. Moreover, the provided data can be incomplete since these services do not scan all ports and may vary port coverage.

\paragraph*{\textbf{External Intelligence Providers}}
Many organizations already collect intelligence on domain names, IP addresses, and related network entities. The usefulness of such information depends on the specific classification task, but for example for malware detection, numerous threat intelligence sources provide lists of known command and control servers. Platforms such as VirusTotal, PhishTank, and Abuse.ch offer valuable security-related data that can be leveraged directly during the labeling process.
However, this approach shares similar drawbacks with active probing. The freshness of the data is often limited, the quality of the intelligence lists strongly affects labeling accuracy, and much of the external intelligence is restricted behind paywalls, which can significantly increase labeling costs. 

\subsubsection{Strategies for Addressing Label Scarcity}
\label{subsub:DataCollection:PracticalStrategies}
Accurately labeling data is often a major challenge. Acquiring labels---especially when relying on external intelligence providers---can be costly and time-consuming. Consequently, several methodologies have been developed to minimize the number of labeled samples required for effective model training.

\paragraph*{\textbf{Semi-supervised approaches}} These methods are designed to work with limited labeled data. Common approaches include self-training~\cite{amini2025self} (where a model's own most confident predictions on unlabeled data are used as new training examples) and label propagation~\cite{iscen2019label} to extend the reach of the initial labels.

\paragraph*{\textbf{Active learning}} This is an interactive approach that is particularly resource-efficient~\cite{settles2009active}. The model identifies the unlabeled flows that it is most \textit{confused} about and presents only those for external labeling---\textit{e.g.}, human expert, external API. This ensures that labeling costs are spent only on the most informative examples.

\subsection{Practical and Operational Challenges}

While the previous sections have detailed the ideal process for generating high-quality flow records, operational deployments are fraught with challenges that can degrade data quality and impact classification performance. This section synthesizes the key challenges that practitioners face.

\paragraph*{\textbf{Network and System Effects}}
The network itself can distort traffic patterns. Congestion and queuing delays add jitter that affects timing features, a phenomenon often called \textit{network effects}. The chosen \textit{observation point} also plays a crucial role: measuring a flow at the client, server, or in the middle of the path can produce very different statistics. In addition, modern \textit{NIC offloading} features such as TCP Segmentation Offload (TSO) on the sender and Generic Receive Offload (GRO) on the receiver can reshape the packet stream. For example, GRO may merge multiple packets into a single large one (up to $\sim$64KB), which can heavily bias packet size statistics if not considered.

\paragraph*{\textbf{Measurement System Artifacts}}
The measurement infrastructure itself introduces noise. High loads can cause packet drops or timestamp inaccuracies. Different export protocols (\textit{e.g.}, NetFlow vs. IPFIX implementations) may have different feature definitions or timestamp resolutions.

\paragraph*{\textbf{Scalability and Performance}}
Modern networks generate traffic at a scale that can overwhelm under-provisioned systems. A flow meter on a core link must handle millions of packets per second and potentially millions of concurrent flows. This necessitates a relentless focus on computational efficiency, often requiring trade-offs between measurement precision and performance, for example, by using statistical sampling (\textit{e.g.}, PSAMP~\cite{rfc5474}) or specialized hardware acceleration.

\paragraph*{\textbf{Concept Drift (Protocol and Application Evolution)}}
The network is not static. Applications are updated, and protocols evolve, causing their behavioral fingerprints to change over time. This phenomenon, known as \textit{concept drift}, can quickly render a trained model obsolete. For example, recent research showed a model's accuracy degrading by 10\% in just one week due to natural traffic drift~\cite{luxemburk2023encrypted}. Operational systems must therefore be designed for continuous monitoring and periodic retraining.

\paragraph*{\textbf{Privacy and Regulatory Constraints}}
Regulations like GDPR now impose strict limits on data collection, processing, and retention. While flow-based methods are less intrusive than DPI, they still require careful design to ensure compliance. This involves implementing robust anonymization techniques (\textit{e.g.}, as guided by RFC 6235~\cite{rfc6235}), defining clear data retention policies, and adhering to the principle of data minimization.

The outlined challenges collectively explain why a model that achieves 99\% accuracy in a clean, controlled lab environment may struggle in a messy, high-scale production network. A successful system is not just one with a clever algorithm, but one whose entire data collection and preparation pipeline is designed to be resilient to these real-world operational realities.

\section{Data Preparation for Machine Learning}
\label{sec:data-preparation}

The previous section described the complex process of collecting network traffic measurements by aggregating packets into flow records. However, such records are rarely suitable for direct use in training ML models due to imperfections, measurement artifacts, or general incompatibility with ML algorithm requirements. This creates the need for a dedicated \emph{data preparation} phase, which is widely recognized as the most critical and labor-intensive step~\cite{Sculley2015}. In this section, we describe the methods commonly employed when designing an ML model based on network flow data, noting that only a subset of these steps is also required during deployment.

\begin{figure}[!t]
	\centering
	\includegraphics[width=1\columnwidth]{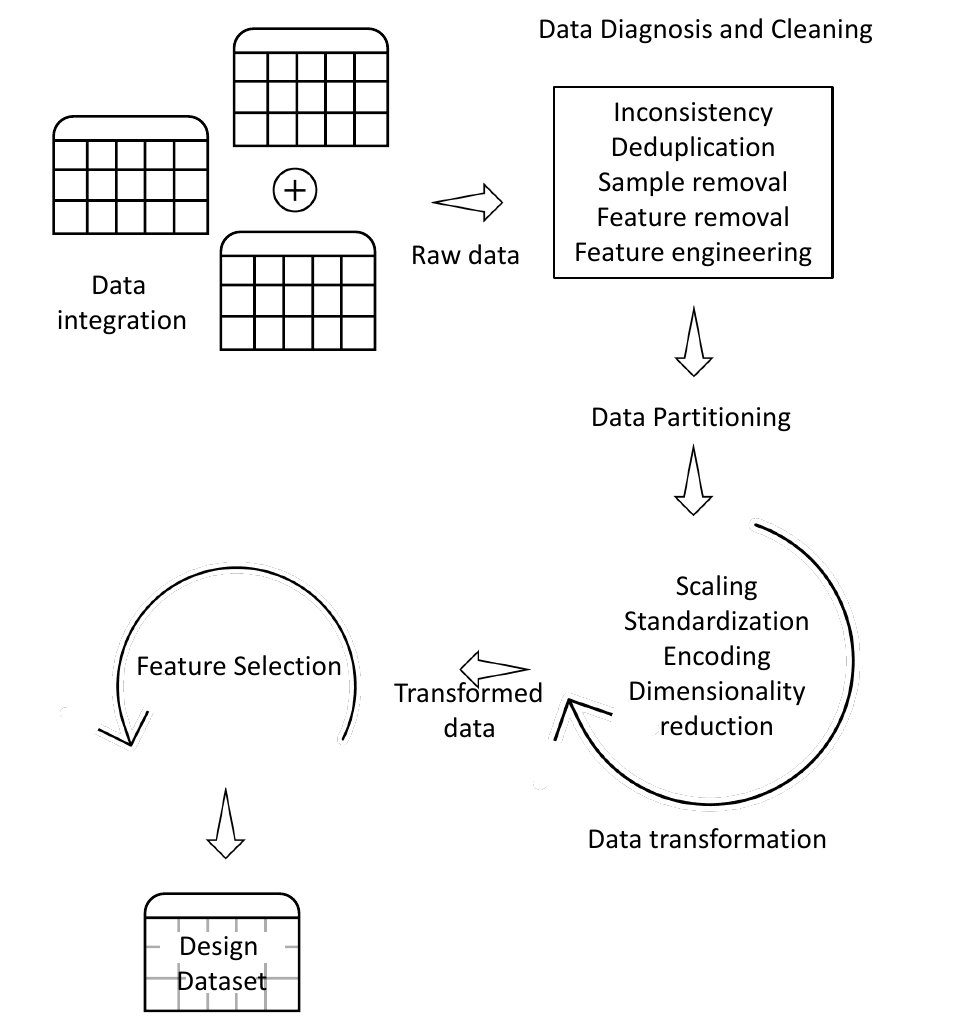}
    \caption{The systematic workflow for data preparation presented in this section.}
	\label{fig:data_preparation}
\end{figure}

\Cref{fig:data_preparation} shows the workflow followed in this section. It begins with \textit{data integration} to unify sources, \textit{cleaning} to correct errors, and \textit{feature engineering} to construct useful predictors.
With a clean and feature-rich dataset in place, the process continues with \textit{data partitioning}, which marks the transition to the \textit{data transformation} phase. This stage involves scaling and encoding features for algorithms, and \textit{dimensionality reduction}---creating a compact and informative feature set through selection or transformation methods.

In practice, data preparation is inherently \textit{iterative}. Later stages often expose the need to revisit earlier steps, for example, insights gained during feature selection may reveal that alternative cleaning is suitable. The workflow accommodates such iteration while maintaining the consistency of experimental protocol and validity of results---data partitioning must remain intact to avoid biased evaluation. This approach is consistent with frameworks like Cross Industry Standard Process for Data Mining (CRISP-DM)~\cite{Schrer2021}.

\subsection{Data Integration}
\label{subsec:data-integration}

Operational networks rarely rely on a single source of truth, instead deploying a mosaic of complementary monitoring technologies, including flow meters on routers, packet capture appliances in critical segments, and application-level monitoring on key servers. Each system observes traffic through a different lens, creating a rich but inconsistent set of perspectives.

\emph{Data integration} is the first step in the preparation workflow, where heterogeneous data streams are combined into a coherent and unified dataset. This process involves resolving differences in measurement formats, timing references, and semantic interpretations. Although often viewed as a technical task, the quality of integration critically affects classification performance: models trained on poorly integrated data may focus on artifacts rather than true traffic patterns~\cite{Putrama2024}. Effective data integration rests on addressing two fundamental challenges: \textit{schema integration} and \textit{entity resolution}.

\subsubsection{Schema Integration}

The first challenge, schema integration~\cite{ElHaddadi2024}, addresses the problem that different monitoring systems use different data structures and definitions to describe traffic. This heterogeneity appears in two primary forms:

\paragraph*{\textbf{Syntactic Heterogeneity}} This arises when identical information is represented in different formats across monitoring systems. For example, one system may record timestamps in microseconds while another uses milliseconds. Similarly, byte counts can be expressed with different units or levels of precision, and even the field names for the same metric may vary---for instance, one system exporting \textit{octetTotalCount} while another reports \textit{bytes}.

\paragraph*{\textbf{Semantic Heterogeneity}} This occurs when the same feature is computed differently across systems. A typical example is \textit{flow duration}: a flow meter may define it using TCP semantics (from the first SYN to the final FIN or RST), whereas a basic router measures only the time between the first and last observed packets, regardless of TCP state. Other semantic discrepancies include differences in flow termination logic, handling of fragmented packets, treatment of retransmissions, and interpretation of protocol flags. Such inconsistencies yield fundamentally different values for ostensibly the same traffic feature, potentially distorting a flow's statistical signature and misleading classifiers.

Resolving both forms of heterogeneity requires establishing a \textit{canonical data model} that provides a single integration target. A standard set of features must be specified with precise semantics (\textit{e.g.}, timestamps expressed in UTC microseconds, traffic volumes measured in bytes), along with system-specific transformation rules that map each data source into this unified format.

\subsubsection{Data Association}

Data association is the process of linking records from different monitoring systems corresponding to the same underlying network activity. For example, a single VoIP call may be observed both by a flow meter on a perimeter router and by a DPI appliance at a security gateway, representing two distinct measurements of the same logical event.

Simple matching based on the 5‑tuple is often insufficient. Records of the same flow captured at different vantage points may exhibit divergent feature values, because of factors such as Network Address Translation (NAT) or partial views caused by asymmetric routing. In addition, timestamps are rarely identical, due to imperfect clock synchronization and the inevitable variations in packet timing as traffic traverses the network.

To address data association properly, algorithms must use more sophisticated techniques than just simple matching~\cite{10.1145/3418896}. In the case of network flow data, this often involves probabilistic methods. The main idea is to compute a \textit{similarity score} between candidate records based on several attributes. Instead of a strict \textit{match/no-match} decision, the system may, for example, estimate a 95\% chance that two records belong to the same session if their 5-tuples are almost identical and their packet counts and timings fall within reasonable bounds. The result of this process is a \textit{composite flow record}, which combines the most reliable information from the associated observations to form an accurate representation of the flow.

\subsubsection{Implementation}

Building a reliable data integration pipeline requires careful attention to several operational details, including temporal alignment through time synchronization protocols~\cite{Levesque2016}, as well as deduplication and resolving conflicts between sources when monitoring devices report different data.  

Moreover, it is necessary to continuously validate the quality of this integration. A common approach is to inject synthetic traffic with known characteristics at multiple points in the network and then verify that the resulting composite records faithfully reproduce the ground truth.  

Although the following sections of this tutorial focus on single-source scenarios for clarity, these integration principles remain essential when extending classification systems from controlled experiments to complex, real-world production environments.

\subsection{Quality Diagnosis and Universal Cleaning}
\label{subsec:initial-cleaning}

Following the integration of disparate data sources into a unified dataset, the workflow proceeds to \textit{data quality diagnosis} and \textit{universal cleaning}. The objective at this stage is to identify and correct unambiguous, structural errors that would otherwise corrupt any downstream analysis. 

It is important to note that certain data quality operations are deliberately postponed until after the dataset has been split into training and testing parts. This is necessary for techniques that rely on learning statistical parameters from the data. Applying such operations to the entire dataset would allow information from the future test set to \textit{leak} into the training process---a phenomenon known as \textit{data leakage}---which can result in overly optimistic performance estimates and poor generalization to unseen data. In this section, we present the \textit{universal} cleaning operations that can be safely applied to the entire dataset, as they rely on fixed logical rules and predefined thresholds, thus forming a methodologically sound practice.

\subsubsection{Exploratory Data Analysis}

The critical first step in cleaning is a targeted investigation of the dataset to reveal its structural integrity and identify obvious flaws~\cite{Tukey1977}. 
This diagnostic check typically involves the following key assessments:

\paragraph*{\textbf{Structural Assessment}} This involves verifying the dataset's dimensions, the data types assigned to each column, and its overall memory footprint. This initial check ensures that the data has been loaded correctly and conforms to the expected format before further analysis.

\paragraph*{\textbf{Completeness Check}} This involves calculating the percentage of missing values for each feature. This is critical for identifying columns that are too sparse to provide reliable information for any subsequent learning task. A feature with a very high proportion of missing data is a primary candidate for removal.

\paragraph*{\textbf{Variance Check}} This involves identifying features that have zero or near-zero variance. Constant or near-constant features provide no discriminative information and unnecessarily increase the dimensionality of the dataset.

\paragraph*{\textbf{Plausibility Check}} This involves using basic summary statistics to find values that violate logical or physical constraints. In network traffic data, this can reveal impossible values such as negative flow durations, packet counts of zero alongside non-zero byte counts, or packet sizes exceeding the MTU.

\paragraph*{\textbf{Redundancy Check}} This involves identifying and counting the number of exact duplicate flow records. These can arise from the data integration process and can unduly influence the training of downstream models if not addressed.

\subsubsection{Applying Universal Cleaning Rules}

Following the diagnosis, the workflow proceeds to the corrective step where the identified issues are systematically addressed. These actions can be considered \textit{universal} as they are based on the logical and structural problems found during the Exploratory Data Analysis.
The usual cleaning actions at this stage include:

\paragraph*{\textbf{Domain-Informed Feature Removal}} Based on expert domain knowledge, features that are known to be irrelevant, unstable, or \textit{leaky} (\textit{i.e.}, contain information that would not be available in a deployment) are manually removed from the dataset.

\paragraph*{\textbf{Useless Feature Removal}} Features identified as having no informational value are dropped. This typically includes columns with an excessively high percentage of missing values (\textit{e.g.}, $> 95\%$) or those with zero or near-zero variance. This action reduces dimensionality and focuses subsequent analysis on more reliable features.

\paragraph*{\textbf{Inconsistency Handling}} Records containing values that are physically or logically impossible are handled. Depending on the nature of the error and the number of affected records, the strategy may involve correcting the value if the fix is obvious (\textit{e.g.}, setting a negative duration to a small positive value if other features are valid) or, more commonly, removing the entire record to avoid propagating measurement errors. This often relies on domain-specific rules, such as invalidating a TCP flow that has a SYN flag but fewer than the three packets required for a minimal handshake~\cite{Kind2009}.

\subsection{Feature Engineering}
\label{subsec:feature-engineering}

Following the quality diagnosis and universal cleaning of the dataset, the workflow proceeds to \textit{feature engineering}. The objective is to derive additional features from raw flow data through domain knowledge and insights from exploratory analysis.

Thorough exploratory analysis typically involves examining individual feature distributions, inter-feature relationships, and temporal patterns, which may reveal valuable predictors for traffic classification. Although ML models are generally expected to infer feature–label relationships automatically, prior knowledge and careful feature preparation often have a substantial impact on performance. The features we engineer can be broadly categorized into two groups: stateless and stateful features.

\subsubsection{Creating Stateless (Per-Flow) Features}

Stateless features are derived solely from the information within a single flow record, without referencing other flows. They can be computed directly from raw records, which makes them suitable for real-time stream processing. Typically, this involves reducing the raw feature set through domain-specific transformations such as \textit{directional aggregation}, \textit{statistical consolidation}, and \textit{protocol-aware engineering}.

\paragraph*{\textbf{Directional Aggregation}} This reduces dimensionality while capturing asymmetry between forward and backward traffic. For example, instead of treating `fwd\_packets` and `bwd\_packets` separately, one can derive `total\_packets` or a ratio feature such as `packet\_ratio`. Such ratios are often highly discriminative, distinguishing balanced flows (\textit{e.g.}, peer-to-peer applications) from asymmetric ones (\textit{e.g.}, video streaming). The same approach applies to byte counts, inter-arrival times, and other directional metrics.

\paragraph*{\textbf{Statistical Consolidation}} This transformation replaces numerous descriptive statistics with the four standard statistical moments, offering a compact yet expressive summary of a feature's distribution within a flow. For metrics such as packet size or inter-arrival time, computing the \textit{mean}, \textit{variance}, \textit{skewness}, and \textit{kurtosis} captures central tendency, dispersion, asymmetry, and tail heaviness, respectively, providing a standardized representation of intra-flow distributions.

\paragraph*{\textbf{Protocol-Aware Engineering}} This method uses domain knowledge to design features that capture specific protocol behaviors. For example, the time ratio of connection establishment time to total flow duration can distinguish short interactive sessions from long bulk transfers~\cite{Bernaille2006}. Other protocol-aware features characterize the density and frequency of traffic bursts, which remain effective even on encrypted flows~\cite{Wang2017}. 

\subsubsection{Creating Stateful (Relational) Features}

Stateful features require context beyond a single flow record. They are derived by aggregating flows grouped by an entity (\textit{e.g.}, source IP or server port) within a time window. By capturing host behaviors and application-level patterns, they reveal discriminative signals not visible at the single-flow level. Although more computationally demanding, their predictive strength often justifies the cost.

\paragraph*{\textbf{Host-Centric Features}} These features describe the behavior of a single client or server over a recent time window, providing a profile of its activity. They are invaluable for anomaly detection as they can quickly reveal behaviors that deviate from normal operation. Common examples include:
\begin{itemize}
    \item The number of concurrent flows originating from a single source IP address.
    \item The rate of new connections initiated by a host (\textit{e.g.}, flows per second).
    \item The number of distinct destination ports or domains contacted by a client in the last minute.
    \item The rate of failed connection attempts (\textit{e.g.}, flows with only a SYN packet and no response).
\end{itemize}
Such features are crucial for detecting malicious activities, including network scanning, command-and-control traffic, and lateral movement within a network.

\paragraph*{\textbf{Service-Centric Features}} These features describe the activity directed toward a specific service (\textit{e.g.}, a web server on port 443). They are particularly useful for monitoring service health and detecting coordinated events like DDoS attacks. Examples include:
\begin{itemize}
    \item The number of unique source IPs connecting to a specific server port over a time window.
    \item The average flow duration or volume for that service.
    \item The geographic diversity of connecting clients, if such information is available.
\end{itemize}

Implementing stateful feature engineering requires careful management of computational and memory overhead, particularly in high-speed networks where large state tables must be maintained. When generating these features for model training on historical data, it is also critical to avoid lookahead bias---a flow at time \textit{t} must be computed solely from flows that occurred strictly before \textit{t}.

\subsection{Data Partitioning}
\label{subsec:data-partitioning}

With a clean, feature-rich dataset prepared, the workflow reaches the critical transition of data partitioning. The goal of this stage is to divide the dataset into disjoint subsets: a training set for model fitting, a validation set for parameter tuning, and a test set for evaluation, which simulates real-world performance on new, unseen data. 

\paragraph*{\textbf{Training Set}} The largest partition, used exclusively to fit the model. All internal parameters (\textit{e.g.}, neural network weights or SVM decision boundaries) are learned from this data.  

\paragraph*{\textbf{Validation Set}} A smaller, separate partition used for model selection and hyperparameter tuning. While the model does not train directly on it, performance on this set guides design choices such as learning rates, tree depth, or architecture, thereby indirectly shaping the final model.  

\paragraph*{\textbf{Test Set}} The final held-out partition, used only once after development is complete. Its role is to provide an unbiased estimate of real-world performance on unseen data, making its integrity critical for reliable evaluation.

The exact proportions of the split depend on the size of the overall dataset; a common starting point is a 60\%-20\%-20\% or 70\%-15\%-15\% split for training, validation, and testing, respectively. For very large datasets, smaller percentages can be allocated to validation and testing while still providing a statistically significant number of samples. 

\subsubsection{Partitioning Strategies: Establishment of the Experimental Protocol}
\label{sec:part-strat}

\begin{figure}[ht]
  \centering
  \begin{subfigure}[b]{0.8\columnwidth}
    \centering
    \includegraphics[width=\linewidth]{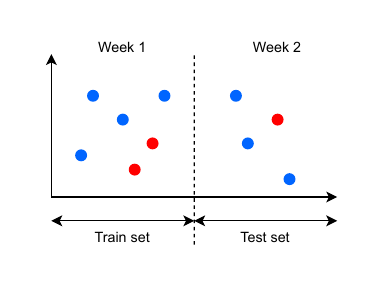}
    \caption{Temporal split: dataset is split based on time.}
    \label{fig:temporal-split}
  \end{subfigure}
  \hfill
  \begin{subfigure}[b]{0.8\columnwidth}
    \centering
    \includegraphics[width=\linewidth]{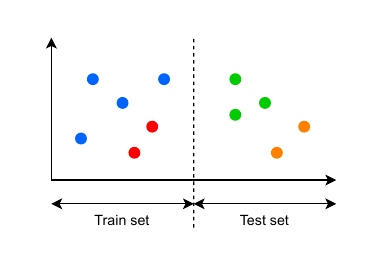}
    \caption{Disjoint split: dataset is split by classes (or origin sources of data).}
    \label{fig:disjoint-split}
  \end{subfigure}
  \caption{Comparison of dataset splitting strategies.}
  \label{fig:splitting-strategies}
\end{figure}

Designing a robust experimental protocol is a critical aspect of ML algorithm development. The specific method of data partitioning depends on the intended use case. In the context of traffic classification, the most commonly adopted experimental protocols include: Random Split, Temporal Split, and Disjoint Split. \Cref{fig:splitting-strategies} compares these strategies schematically. Moreover, in this section, we also describe the Out-of-Distribution split.

\paragraph*{\textbf{Random Split}} This split is the most commonly used method for training an ML model. The data are divided into three sets---train, validation, and test sets---which are assigned randomly using different sampling algorithms. Various sampling algorithms can be employed to perform this split, each designed to preserve specific characteristics of the data. Among these, stratified sampling is widely used, which maintains the original distribution of class labels (when available) across all subsets. This property is especially important when dealing with imbalanced datasets and rare classes.

\paragraph*{\textbf{Temporal Split}} This split is explicitly designed to preserve the chronological order of the data~\cite{luxemburk2023datazoo,malekghaini2022data,luxemburk2023encrypted}. In this approach, all samples in the training set are captured from an earlier time period than those in the test set, as illustrated schematically in~\Cref{fig:temporal-split}. The motivation behind this evaluation protocol lies in the assumption that network traffic evolves over time. This approach tests the model's ability to generalize to future traffic patterns and network changes, mirroring real-world deployment conditions.

\paragraph*{\textbf{Disjoint Split}} This split prevents overlap between training and test sets by ensuring that no duplicate or highly similar samples occur across them---a common pitfall when identical traffic patterns appear multiple times in network data~\cite{luxemburk2021detection}. As shown in~\Cref{fig:disjoint-split}, this split enforces a separation of network endpoints, providing a stricter test of generalization. The need arises from the deterministic nature of traffic: for example, repeated connections from the same device to a specific API often produce nearly identical traces. By contrast, connections from different devices to the same API can vary due to differences in TCP stack behavior, network topology, or wireless conditions. Partitioning can also be applied at the application level: for instance, in application-level classification tasks such as video streaming, one vendor’s traffic (\textit{e.g.}, YouTube) may be included in training while another's (\textit{e.g.}, Vimeo) is reserved for testing. Such splits encourage models to capture features generalizable across providers offering similar services. 

\paragraph*{\textbf{Out-of-distribution (OOD) Split}} This is a specific case of the disjoint split, where the test set contains samples from categories unseen during training~\cite{Yang2021}. Instead of forcing the classifier to assign these samples to known classes, the evaluation expects it to reject them, thereby testing its rejection capability---an essential requirement given the dynamic nature of network traffic and the constant emergence of new services. 

\subsubsection{k-fold cross-validation}
Although a single train–validation split is often sufficient, \textbf{k-fold cross-validation} provides a more robust approach for assessing model performance during development. It is important to emphasize that the k-fold cross-validation is used solely to separate training and validation subsets. The test data must remain entirely excluded from this procedure; therefore, the test set should be defined and isolated before performing cross-validation.

\begin{figure}[t]
  \centering
  \includegraphics[width=\columnwidth]{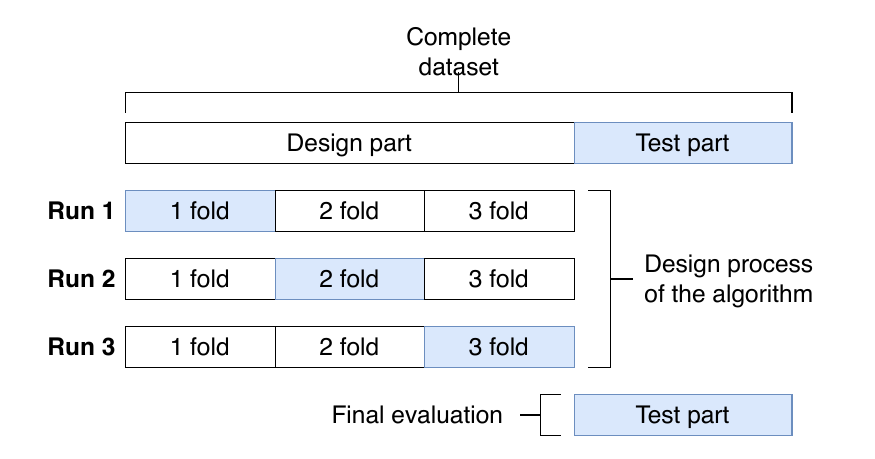}
  \caption{The k-fold cross-validation process (with $k=3$). The training data are split into 3 folds. The process iterates 3 times, with each fold serving as the validation set once, providing a robust estimate of model performance.}
  \label{fig:kfold}
\end{figure}

As illustrated in \Cref{fig:kfold}, the process is as follows:
\begin{enumerate}
    \item The dataset is split into design and test parts.
    \item The design part is split into \textit{k} equal-sized, non-overlapping subsets, or \textit{folds} (\textit{e.g.}, $k=3$ or $k=5$).
    \item The model is trained \textit{k} times. In each iteration, one fold is held out as a temporary validation set, and the model is trained on the remaining $k-1$ folds.
    \item The performance obtained from each of the $k$ iterations is averaged to produce a single estimate. In addition to the mean score, it is often useful to report the standard deviation across folds to assess the stability of the performance values.
\end{enumerate}

The partitioning into folds can follow the strategies described in~\Cref{sec:part-strat}. However, standard random k-fold cross-validation is generally incompatible with temporal splits because it breaks chronological order across validation folds. Time-aware cross-validation variants exist, but here we focus on preserving temporal consistency at least between the design and test parts.

Removing temporal validation consistency is not the only disadvantage. The approach also comes with significantly higher computational costs. The process of designing an ML algorithm involves repeated cycles of training and performance measurement. With k-fold cross-validation, these cycles are replicated $k$ times, resulting in $k \times$ more training runs. When a single training session requires many hours to complete, the added computational demand can prevent its use.

\subsubsection{Preserving Evaluation Integrity and Preventing Data Leakage}

The key rule for preserving evaluation integrity is simple: \textbf{partition data before learning any data-dependent parameters and forget the existence of the test set.} Violating this order leads to subtle but serious problems. For example, if features are scaled to $[0,1]$ using min--max values computed from the whole dataset before splitting, the scaler has already incorporated information from the test set. In practice, however, the test set must be scaled using only statistics from past data, so the training and inference distributions no longer match, degrading performance.

This methodological error is called \textit{data leakage}---test set information influencing training---which breaks the assumption that the test set represents unseen future data and often results in overestimated performance. To avoid this, \textbf{all transformations in the following sections must be applied on already-partitioned data}, with parameters derived exclusively from the training set.

\subsection{Data Transformation and Parameterized Cleaning}
\label{subsec:data-transformation}

With the data partitioned, the workflow proceeds to \textit{data transformation}, which converts engineered features into formats suitable for ML algorithms. Data transformation addresses the scale differences, categorical values, and skewed distributions. This stage must strongly follow the \textit{fit/transform paradigm}, where each statistical data modifier is \textit{fitted} only on the \textit{training data} and then applied to remaining partitions (see~\Cref{subsec:data-partitioning}).

\subsubsection{Scaling Numerical Features}

This transformation handles large scale disparities among numerical features. In network traffic data, values can range from byte counts in the millions to inter-arrival times in microseconds, each with different min/max values. Algorithms relying on distance metrics (\textit{e.g.}, $k$-NN) or gradient-based optimization (\textit{e.g.}, neural networks) are highly sensitive to such differences, with large-range features dominating and reducing the impact of smaller-range ones. Scaling places all features on a common scale (usually [0, 1]), ensuring balanced contributions of all features in the classification process.

Several scaling methods exist, each with specific properties and trade-offs. The choice of scaler should be informed by the characteristics of the data, particularly its sensitivity to outliers.

\paragraph*{\textbf{Standardization (Z-score Scaling)}} This method rescales features to a mean of 0 and a standard deviation of 1, defined as:
\begin{equation*}
x'_{standard} = \frac{x - \mu}{\sigma}
\end{equation*}
where $\mu$ and $\sigma$ denote the feature’s mean and standard deviation. Standardization is a common default, especially for algorithms like SVMs and linear models, but its dependence on the mean and standard deviation makes it highly sensitive to outliers---frequent in long-tailed network feature distributions (\textit{e.g.}, flow duration).

\paragraph*{\textbf{Min-Max Scaling}} This method rescales features to a fixed range, typically [0, 1], using:
\begin{equation*}
x'_{MinMax} = \frac{x - \text{min}(x)}{\text{max}(x) - \text{min}(x)}
\end{equation*}
It is commonly used in neural networks, which require bounded inputs. However, it is highly sensitive to outliers---an extreme value can distort the scaling by compressing most data into a narrow portion of the [0, 1] range, reducing feature separability.

\paragraph*{\textbf{Robust Scaling}} This method resists outliers by using quantile-based statistics:
\begin{equation*}
x'_{Robust} = \frac{x - Q_2(x)}{\text{IQR}(x)}
\end{equation*}
where $Q_2$ is the median and IQR (interquartile range) = $Q_3 - Q_1$. Centering on the median and scaling by the central 50\% of values minimizes the influence of extreme data points. For skewed network traffic features, robust scaling is often the most suitable choice.

\paragraph*{\textbf{A Note on Terminology: Scaling vs. Normalization}}
It is important to clarify the terminology used here. \textit{Scaling} refers to changing the \textit{range} of the data while preserving the \textit{shape} of its distribution. The methods discussed above---Standardization, Min-Max, and Robust Scaling---are all scaling techniques. In contrast, \textit{Normalization} (or more accurately, distribution reshaping) refers to changing the fundamental \textit{shape} of the distribution, often to make it more Gaussian. This second task is discussed in the advanced section on reshaping distributions.

\subsubsection{Encoding Categorical Features}

While most flow features are numerical, network data also includes categorical attributes representing discrete values such as `Protocol` (such as TCP, UDP, ICMP) and TCP `Flags` (such as SYN, ACK, and FIN). Because many ML models require numerical input, these features must be transformed into numeric form. The appropriate encoding method depends primarily on the feature's cardinality---the number of distinct categories it contains.

\paragraph*{\textbf{Low-Cardinality Features}} have a few unique values, such as `L4 Protocol`. In this case, the preferred encoding method is \textit{One-Hot Encoding}. This creates a binary feature for each category. For example, [TCP, UDP, ICMP] becomes `is\_TCP`, `is\_UDP`, and `is\_ICMP`, with a TCP flow represented as `[1, 0, 0]`. This approach prevents introducing false ordinal relationships between categories, which could otherwise mislead certain models.

\paragraph*{\textbf{High-Cardinality Features}} have thousands of unique values (such as IP addresses or ports), making one-hot encoding impractical due to excessive dimensionality. A practical alternative is \textit{domain-aware binning}, which groups categories into meaningful ranges---for instance, port numbers can be categorized as `Well-Known (0–1023),` `Registered (1024–49151),` and `Dynamic/Private (49152–65535).` This approach preserves semantic context while greatly reducing dimensionality, and the resulting bins can then be one-hot encoded. More advanced techniques, such as \textit{target-based encoding}~\cite{pargent2022regularized} or \textit{embedding models}~\cite{corizzo2020feature}, are also available but demand considerably more effort to implement and tune effectively.

\subsection{Supervised Feature Selection}
\label{subsec:supervised-feature-selection}

After transforming the data into a suitable format, the workflow advances to supervised feature selection. Unlike feature engineering, which generates a broad set of candidates, this step reduces dimensionality by retaining only the most predictive features. It mitigates the \textit{curse of dimensionality}, reducing computational cost and overfitting risk. We must stress that the feature selection process must be conducted exclusively on training and validation data, and never on the test set (see~\Cref{subsec:data-partitioning}).

\subsubsection{The Rationale for Dimensionality Reduction in Traffic Analysis}

The \textit{curse of dimensionality} is a real, practical challenge that arises in high-dimensional analysis. It is particularly severe in network traffic analysis, where combinatorial feature engineering can produce hundreds of features. As dimensionality grows, the feature space volume increases exponentially, making data increasingly \textit{sparse}. Maintaining density would require exponentially more flow examples, which is rarely feasible. Sparse data increases the risk of models learning noise rather than true traffic patterns, hurting generalization.

Moreover, algorithms such as $k$-Nearest Neighbors depend on distance metrics. In high dimensions, distances between points tend to converge~\cite{Beyer1999}, making it hard to separate meaningful differences since all points appear similarly distant.

Finally, more features increase the \textit{computational complexity} of learning algorithms. Higher dimensionality means more parameters, slowing training and inference and raising the risk of \textit{overfitting}, where the model memorizes training noise instead of capturing the underlying signal---leading to poor performance on unseen data.

\subsubsection{Dimensionality Reduction via Supervised Feature Selection}

With the rationale for dimensionality reduction established, the selection of an appropriate methodology is the next step. These feature selection techniques are typically categorized into three families---\textit{Filter}, \textit{Wrapper}, and \textit{Embedded methods}.

\paragraph*{\textbf{Filter Methods}}
These methods rank features by their statistical relationship to the target variable, serving as a model-agnostic pre-processing step. They are highly efficient and provide a quick way to reduce large feature spaces. Common techniques include univariate statistical tests, such as the \textit{Chi-squared} test for categorical features and \textit{Mutual Information} for capturing both linear and non-linear dependencies. Features are ranked by their scores, and the top-$N$ or top percentage are retained. Although filter methods are fast, they assess each feature independently and overlook inter-feature interactions and redundancies.

\paragraph*{\textbf{Wrapper Methods}}
These methods \textit{wrap} a specific ML model, using its validation performance to evaluate different feature subsets. Feature selection is framed as an optimization problem---individual feature subsets are tested, and the one yielding the best model performance is chosen. A classic approach is \textit{Sequential Feature Selection}, which works via \textit{forward selection} (adding features that most improve performance) or \textit{backward elimination} (removing those that least degrade it). While wrappers can produce model-optimized feature sets, they are computationally expensive since each feature subset requires model training and evaluation.

For large feature spaces where sequential search is impractical, heuristic methods such as \textit{Genetic Algorithms} can efficiently explore complex feature combinations and uncover high-performing subsets~\cite{Halim2021}. Their main limitation, however, remains the high computational cost.

\paragraph*{\textbf{Embedded Methods}}
These methods integrate feature selection directly into model training, balancing the efficiency of filters with the tailored performance of wrappers. The selection logic is built into the learning algorithm itself. Models with \textit{L1 regularization} (\textit{e.g.}, Lasso regression) penalize coefficients and can shrink those of irrelevant features to zero, effectively removing them. Tree-based ensembles such as \textit{Random Forest} also provide embedded selection: during training, features are evaluated at each split for their discriminative power, and this is aggregated into robust \textit{feature importance} scores that guide the selection of the most influential subset. Although embedded methods handle feature selection internally, they can still benefit from preliminary filtering. By removing clearly irrelevant features early, we reduce computational overhead for subsequent retraining and experiments.

\subsubsection{Dimensionality Reduction via Feature Transformation}
\label{subsec:feature-transformation-reduction}

The preceding section detailed how to reduce dimensionality by selecting the most informative subset of the \textit{original} features. This section presents an alternative and complementary philosophy: reducing dimensionality by \textit{transforming} the data into an entirely new, smaller set of \textit{artificial} features---essentially performing a lossy compression. Such transformations are also useful for visualization, as high-dimensional data are difficult to plot directly. There are three main approaches: linear transformations, non-linear transformations, and deep learning-based feature embeddings. 

\paragraph*{\textbf{Linear Transformations}}
\begin{figure}[t]
	\centering
	\includegraphics[scale=.5]{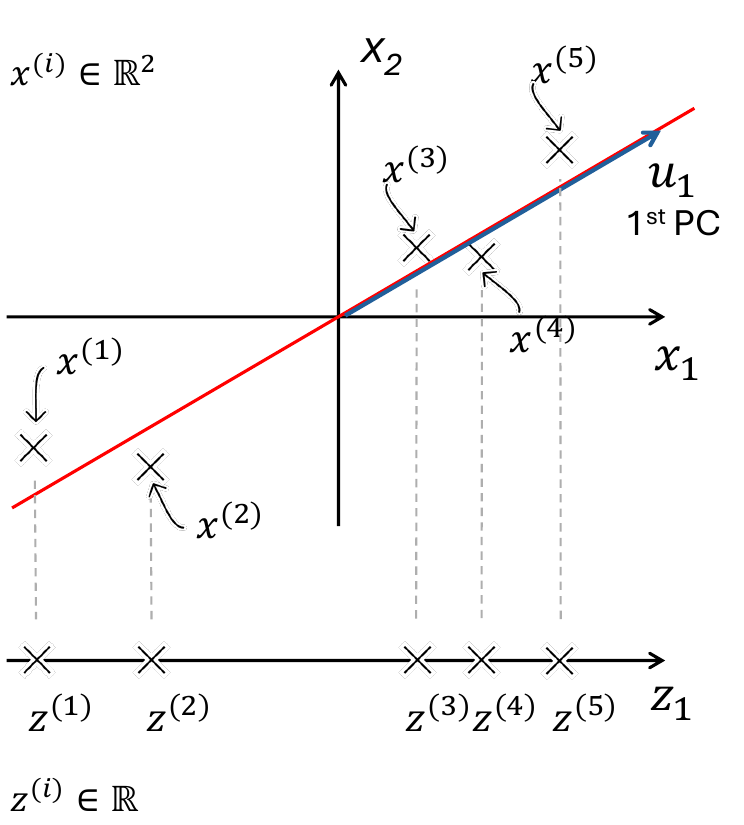}
	\caption{PCA finds the direction of maximum variance $\mathbf{u}_1$ (the first principal component) and projects the data points $\mathbf{x}^{(i)} \in \mathbb{R}^2$ onto this axis via $z^{(i)} = \mathbf{u}_1^T \mathbf{x}^{(i)}$, reducing dimensionality from $\mathbb{R}^2$ to $\mathbb{R}$ while preserving the most important patterns in the data.}
	\label{fig:pca}
\end{figure}

While numerous linear transformations exist, Principal Component Analysis (PCA) is the most prevalent technique for dimensionality reduction. It finds orthogonal axes (principal components) ordered by their ability to capture maximum data variance. The first component captures the direction of greatest variation, the second captures the next highest orthogonal variance, and so forth. This process de-correlates features and compresses information from many original features into fewer new ones. PCA's operational principle is illustrated in~\Cref{fig:pca}.

PCA is suitable for network traffic data due to strong correlations among flow features (\textit{e.g.}, packet counts, byte counts, and flow duration). By identifying these redundancies, PCA can create compact representations in which a relatively small number of principal components captures a large fraction of the variance; the exact number depends on the dataset and analysis objective~\cite{Jolliffe2016}. Though its linearity limits its ability to capture complex non-linear traffic patterns, its simplicity and efficiency make it an excellent initial choice for dimensionality reduction.

\paragraph*{\textbf{Non-Linear Transformation}}
When relationships among traffic features are highly complex, advanced methods such as t-Distributed Stochastic Neighbor Embedding (t-SNE)~\cite{vandermaaten2008visualizing} and Uniform Manifold Approximation and Projection (UMAP)~\cite{mcinnes2018umap} can help. These algorithms preserve local neighborhood structures, ensuring that flows close in the original high-dimensional space remain close in the reduced representation. Their projections typically produce 2D or 3D spaces, making them well suited for visualization and exploratory analysis. Using these embeddings as inputs for classification models requires caution: the transformation must be fit on training data only and applied consistently to validation and test data, and the resulting embeddings can be sensitive to hyperparameters and implementation details.

\paragraph*{\textbf{Deep Learning-Based Feature Embeddings}}

A state-of-the-art alternative to the previously described methods is the use of \textit{Deep Learning-Based Feature Embeddings}. These are obtained through self-supervised training of a neural network, which is tasked with learning compressed representations. The resulting vectors represent features in a novel embedding space, with the number of features determined by the dimensionality of this space.

\begin{figure}[t]
	\centering
	\includegraphics[scale=.45]{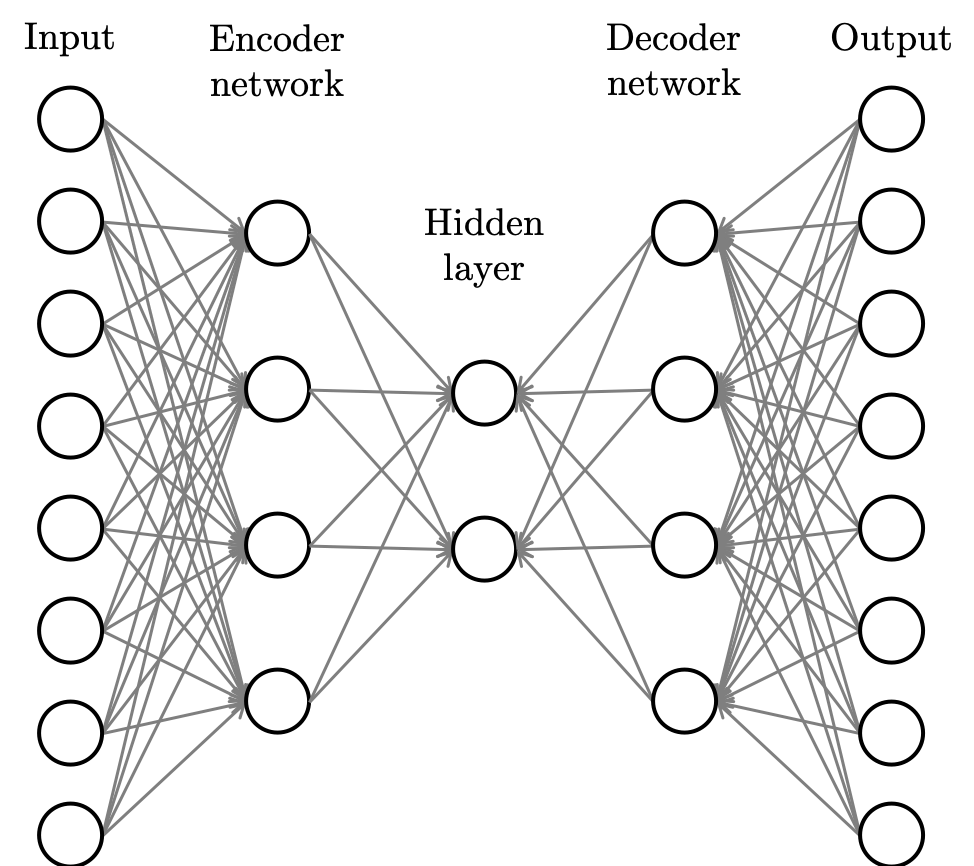}
	\caption{Autoencoder: the encoder network maps the original, high-dimensional data into its compressed, low-dimensional representation (hidden layer), and then, the decoder network maps it back to a high-dimensional representation (reconstructed data).}
	\label{fig:ae}
\end{figure}

A typical example is the \textit{Deep Autoencoder} (shown in~\Cref{fig:ae}), which consists of two components: an encoder that compresses high-dimensional flow data into a compact latent representation (\textit{e.g.}, 16–64 dimensions), and a decoder that reconstructs the original data from this compressed form. The training objective is to minimize the reconstruction error, forcing the encoder to capture the most informative characteristics of the traffic. For feature embedding applications, only the encoder is used, while the decoder serves solely for training.

\subsection{Training Set Preprocessing}
\label{subsec:advanced-training-prep}

Since many algorithms, such as $k$-NN or SVM, depend heavily on the training data and lack internal data-adaptation mechanisms, it is beneficial to preprocess the label and sample distributions within the training partition. However, the operations described in this section must be applied carefully, as they fundamentally modify the training dataset. Moreover, they must never be used on the validation or test partitions---the goal is to adjust the training data in a way that improves performance on original and unseen validation and test sets.

\subsubsection{Class Imbalance}
In many network security and management tasks, such as intrusion detection, the prevalence ratio between benign and malicious samples is often highly imbalanced, commonly reaching 100:1 or even 1000:1. Under such extreme imbalance, some ML algorithms tend to ignore the minority class, resulting in an ineffective detector. To mitigate this, undersampling and oversampling techniques can be applied to reduce the severity of the imbalance. In practice, both techniques are often used together.

\paragraph*{\textbf{Undersampling}} methods reduce the size of the majority class (\textit{e.g.}, benign traffic), for instance by randomly removing benign flows until a more balanced ratio between majority and minority classes is achieved. More sophisticated variants may remove near-duplicate samples or replace dense clusters of flows with the same label by their corresponding centroids.  

\paragraph*{\textbf{Oversampling}} methods increase the representation of the minority class. A naive approach is to simply duplicate existing rare instances, but such a method can lead to overfitting. More advanced approaches synthetically generate new, plausible instances for the minority class. The two most commonly used algorithms are Synthetic Minority Oversampling Technique (SMOTE)~\cite{Chawla2002} and Adaptive Synthetic Sampling (ADASYN)~\cite{ea2008adaptive}. SMOTE generates new samples uniformly across the minority class---potentially leading to oversampling in already dense regions, whereas ADASYN addresses this issue by focusing on hard-to-classify instances located in sparse areas or near decision boundaries. Usually, both algorithms are evaluated experimentally, and the one yielding better performance is selected.

\subsubsection{Outlier Removal}

Despite its similarity to the prior removal of impossible or inconsistent values (see~\Cref{subsec:initial-cleaning}), this step targets statistically rare yet potentially valid observations whose extreme magnitudes can unduly influence training, even after robust scaling. A detection rule is applied on the training data to flag rows with extreme values, and those rows are then removed. Removing such outliers often comes with a trade-off: it may improve generalization by preventing overfitting to corner cases, but it also risks discarding genuinely informative events. The two primary approaches are:

\paragraph*{\textbf{Z-Score}} This method defines outliers as data points that fall a certain number of standard deviations (typically 3 or more) away from the mean. While straightforward, its effectiveness is limited for the skewed distributions common in network traffic, as the mean and standard deviation are themselves heavily influenced by the very outliers the method aims to detect.

\paragraph*{\textbf{Interquartile Range (IQR)}} This method is a more robust method that is resistant to the influence of extreme values. It defines outliers as points falling outside a range defined by the quartiles of the data (typically below $Q_1 - 1.5 \times \text{IQR}$ or above $Q_3 + 1.5 \times \text{IQR}$). Because it relies on the median and quartiles, it provides a much more reliable measure of statistical rarity for non-Gaussian distributions, making it the preferred method for most network traffic features.

\subsection{Emerging Challenges in Data Preparation}
\label{subsec:emerging-challenges}

The systematic approach to data preparation provided in this section offers a solid foundation for handling network traffic. Nevertheless, the field is constantly evolving, and the selected approaches need to adapt to emerging trends in the field of computer networks, novel regulations or emerging attacks on AI systems. 

\paragraph*{\textbf{Increasing Homogenization}}
A major trend is the growing uniformity of traffic at the transport layer, caused by the widespread adoption of standardized protocols such as HTTP/3 (over QUIC)~\cite{rfc9114} and TLS 1.3~\cite{rfc8446}. This homogenization causes traffic from fundamentally different applications to appear nearly identical at the packet level, undermining the discriminative power of many traditional stateless features. Consequently, the importance of stateful (relational) features is increasing, as higher-level behavioral signatures and inter-flow relationships emerge as the primary remaining sources of discriminative signal.

\paragraph*{\textbf{Privacy-Preserving Analysis}}
Cross-domain privacy regulations such as GDPR have made the \textit{collect everything} approach obsolete, which can significantly shift the methods used in the initial stage of our pipeline---Data Integration (see~\Cref{subsec:data-integration}). 
The pipelines should be prepared to incorporate privacy-by-design as a fundamental principle. This can involve embedding anonymization techniques directly into the collection and integration phases, or adopting advanced approaches such as \textit{differential privacy}~\cite{Yang2024}. Nevertheless, these techniques often come at the cost of reduced performance and may reduce model utility, and therefore require careful evaluation of the privacy--utility trade-off for the target deployment.

\paragraph*{\textbf{Adversarial Attacks}}
As ML becomes integral to security systems, adversaries are actively crafting evasion strategies that specifically exploit weaknesses across the data preparation pipeline, not only the classifier itself. Malicious traffic can be engineered to mimic the statistical profile of benign applications, bypassing detectors that rely on these behavioral patterns. Apart from maximizing the accuracy, future data preparation pipelines should also prioritize adversarial robustness and make data pipelines resilient to manipulation.

\paragraph*{\textbf{Lack of Standard Feature Set}}
Even though ML algorithms are widely applied in traffic analysis use cases, there is still no standardized feature set that would enter the Data Preparation phase. Flow features vary significantly from paper to paper, often with notable limitations. Some features are impractical for real-world deployment because their computation is too resource-intensive for real-time processing, or because the necessary input data are simply not available in operational environments. This heterogeneity slows progress by fragmenting datasets and evaluations, making cross-paper benchmarking tedious due to feature alignment or conversion steps. Future systems should adopt one or more standard input feature sets to enable fair evaluation, reproducibility, and cumulative advances across the field.

\paragraph*{\textbf{Data Drift}}
In addition, computer networks evolve continuously: new protocols and versions appear, devices are added or retired, and bandwidth capacities grow. These dynamics make networks highly variable environments in which the statistical properties of traffic shift over time (data drift), often degrading the performance of ML models. Consequently, future data preparation pipelines should be highly automated, supporting continuous feature selection and adaptation, and leveraging techniques such as AI-driven agents to account for changing feature importance and other distributional shifts.

\section{Machine Learning Model Development}
\label{sec:machine-learning}

Machine learning enables systems to identify structural patterns in data and improve their performance without being explicitly programmed. Its applications range from recommendation systems to intelligent agents trained via reinforcement learning to act in the physical world. 

This tutorial, however, focuses on network traffic classification. Algorithms in this domain typically fall into two categories: supervised and unsupervised learning. Supervised learning aims to derive a function that maps inputs to outputs, using datasets containing both input features and corresponding output labels~\cite{10.5555/2588013}. In contrast, unsupervised learning aims to uncover structure in data without predefined output labels. In the following text, we focus solely on the supervised classification approaches, since unsupervised methods typically require complex post-processing and analysis steps to infer meaningful traffic categories, which are beyond the scope of this tutorial.

Supervised learning is suitable for classification (\textit{e.g.}, determining what category a network traffic flow belongs to) and regression (\textit{e.g.}, predicting the volume of transmitted data in bytes for a given future date) problems. The system is trained to detect the underlying patterns and relationships, enabling it to yield good results when presented with new, never-before-seen input data.

Throughout this tutorial, we will use supervised learning to classify network flows into application types as an example use case for traffic classification. When learning to classify the application types, the learning algorithm takes thousands of network traffic flows with labels containing the correct application type each flow represents. The algorithm first learns the relationship between the flows and their associated types. Then, it applies that learned relationship to classify completely new flows (without labels) that the machine has not seen before.

\subsection{Introduction to Supervised Classification}
\label{sec:ml-supervised-classification}

Formally expressed, the classification ML algorithm aims to approximate a projection $f: X\rightarrow Y$, where $Y$ is a finite set of unique labels and $X$ is the vector space of input features. When applied to the traffic type classification, we will get:

\begin{equation*}
	y = f(x),
\end{equation*}

\noindent where

\begin{conditions*}
 x \in X  & input vector of flow features \\
 y \in Y  & output, application type of the flow (label)
\end{conditions*}

As mentioned in~\Cref{subsec:data-partitioning}, the function is designed on data from the training set and then evaluated on the testing set. 

\subsection{Performance Measurement}
Performance evaluation in ML is a wide discipline including numerous specialized metrics, each designed to target a specific objective. In this section, we focus primarily on binary classification metrics, which will be subsequently generalized to the multiclass scenarios. For that purpose, we will simplify the example use case: Video-streaming identification---the classes are thus \textit{video streaming} vs \textit{other traffic}.

In binary classification, where only two classes are considered, the predicted labels (outputs) can generally be divided into the following categories:

\begin{LaTeXdescription}
\item[True Positives (TP)] are the positive cases where the classifier correctly identified them. For example, if the actual class value indicates that a flow is of a particular type, and the predicted class tells the same.

\item[False Positives (FP)] are the negative cases where the classifier incorrectly identified them as positive. For example, if the actual class value indicates that a flow does not belong to a particular class, the predicted class says it belongs.

\item[True Negatives (TN)] are the negative cases where the classifier correctly identified them. For example, if the actual class indicates that a particular flow is not of a specific type, and the predicted class tells the same thing.

\item[False Negatives (FN)] are the positive cases where the classifier incorrectly identified them as negative. For example, if the actual class value indicates that a flow is of a certain type and the predicted class tells that it is not.
\end{LaTeXdescription}

\begin{table}[ht]
    \centering
    \renewcommand{\arraystretch}{1.19}
    \caption{The basic framework of the confusion matrix.}
    \label{tbl:confusion}
    \begin{tabular}{cc|c|c|}
        \multicolumn{1}{c}{} & \multicolumn{1}{c}{} & \multicolumn{2}{c}{\textbf{Predicted Class}} \\
        \multicolumn{1}{c}{} & \multicolumn{1}{c}{} & \multicolumn{1}{c}{\emph{P}} & \multicolumn{1}{c}{\emph{N}} \\
        \cline{3-4}
        \multirow{6}{*}{\rotatebox[origin=c]{90}{\textbf{Actual Class}}} 
            & \multirow{3}{*}{\rotatebox[origin=c]{90}{\emph{P}}} 
                & True & False \\ 
            & & Positives & Negatives \\
            & & (TP) & (FN) \\
        \cline{3-4}
            & \multirow{3}{*}{\rotatebox[origin=c]{90}{\emph{N}}} 
                & False & True \\ 
            & & Positives & Negatives \\ 
            & & (FP) & (TN) \\
        \cline{3-4}
    \end{tabular}
\end{table}

The occurrences or proportions of TP, FP, TN, and FN can be organized into a matrix, as shown in~\Cref{tbl:confusion}, referred to as the confusion matrix. The use of such matrices is extremely common in binary classification, as they capture complete information about classification outcomes. However, because a confusion matrix contains four separate values, it is often impractical for direct comparison between algorithms. Consequently, various performance metrics have been developed that combine the values from the confusion matrix into a single numerical measure. The most common metrics are:

\paragraph*{\textbf{Accuracy}} This metric is the ratio of correctly predicted observations to the total observations:
    \begin{equation*}
        Accuracy = \frac{TP + TN}{TP + FP + FN + TN}
    \end{equation*}
    
It is one of the simplest and most intuitive performance metrics, and is therefore widely used in research studies. However, in highly imbalanced scenarios it can present a misleading view, as it does not account for the prevalence of each class. For example, in a dataset with a class imbalance ratio of 99:1, a classifier that predicts only the majority class label---despite being completely ineffective---would still achieve an accuracy of 0.99. A balanced version of accuracy also exists, in which the contribution of each class is weighted according to its prevalence. While balanced accuracy is straightforward to compute, it can be less intuitive to interpret than standard accuracy, thereby reducing the simplicity advantage of standard accuracy.
    
\paragraph*{\textbf{Precision}} This metric is the ratio of correctly predicted positive observations to the total predicted positive observations: 
    
\begin{equation*}
    Precision =  \frac{TP}{TP + FP}
\end{equation*}
    
Precision provides information about a classifier's performance concerning false positives. In our example classification use case, it provides the answer to the following question: \textit{How many samples identified as video-streaming are actually video streaming?} The metric does not consider the number of actually positive samples missed by classifier. Even if only one sample of a certain class has been captured correctly, precision will be 100\%. Precision is useful when the goal is to minimize false positives. Then, precision should be as close to 100\% as possible.  
    
\paragraph*{\textbf{Recall}} This metric is also known as sensitivity or true positive rate (TPR). It is the ratio of correctly predicted positive samples to all positives samples:

\begin{equation*}
    Recall = \frac{TP}{TP + FN}
\end{equation*}
    
Recall provides information about a classifier’s performance with respect to false negatives. In our example use case, it answers the question: \textit{How many video-streaming samples have been successfully recognized?} It is therefore complementary to Precision, and these two metrics are typically used together. Recall is particularly useful when non-uniform class sampling is employed---for instance, if video-streaming traffic is represented in the dataset at a different ratio than other traffic types. Because Recall does not depend on metrics involving the opposite class, it is unaffected by these sampling conditions.

\paragraph*{\textbf{$\text{F}_\beta$-measure}} This metric is a weighted harmonic mean of Precision and Recall. 
\begin{equation*}
    \text{F}_{\beta} = (1 + \beta^2) \cdot \frac{\text{Precision} \cdot \text{Recall}}{(\beta^2 \cdot \text{Precision}) + \text{Recall}}
\end{equation*}
The parameter $\beta$ determines the relative weight given to Recall: higher values of $\beta$ prioritize Recall over Precision, with the influence scaling as $\beta^2$. For instance, $\beta=2$ weights Recall four times more than Precision. The choice of $\beta$ requires careful consideration of the application context and operational requirements. In practice, the most common case is $\beta=1$, meaning Recall and Precision are equally important, forming the $\text{F}_1$-score:
    
\begin{equation*}
    \text{F}_{1}\text{-score} = 2 \cdot \frac{\text{Precision} \cdot \text{Recall}}{\text{Precision} + \text{Recall}}
\end{equation*}
    
The $\text{F}_{1}$-score equally weights Precision (which penalizes false positives) and Recall (which penalizes false negatives), making it more robust to class imbalance than accuracy. Nevertheless, it is advisable to evaluate models using multiple metrics to obtain a comprehensive understanding of their performance, rather than relying solely on a single measure.

\paragraph*{\textbf{Area Under the Curve}} This is another metric considering both TP and FP~\cite{FAWCETT2006861}. It uses the relationship of the true positive rate (TPR) and the false positive rate (FPR) to establish the performance of the classifiers. The TPR is equivalent to Recall as defined above, while the FPR represents the proportion of false positives relative to all actual negatives:

\begin{equation*}
    FPR = \frac{FP}{FP+TN}
\end{equation*}

\begin{figure}[ht]
	\centering
	\includegraphics[width=.6\columnwidth]{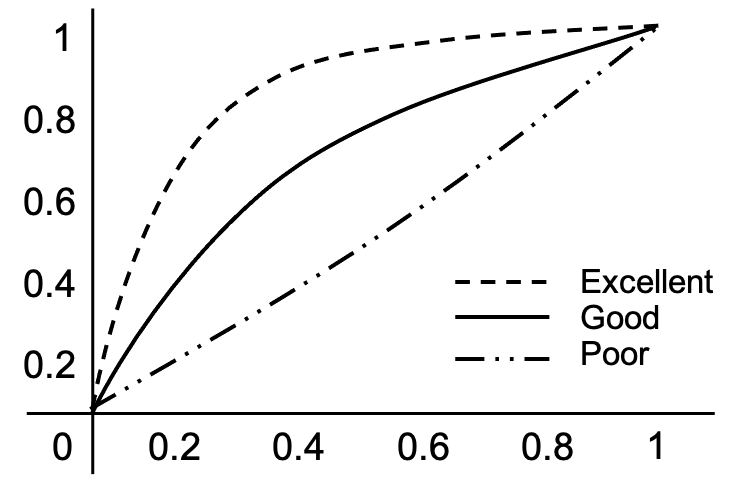}
	\caption{Comparing ROC curves.}
	\label{fig:roc}
\end{figure}

The classification models often produce class scores (often interpretable as probabilities) rather than labels. Therefore, a decision threshold is required to convert scores into binary predictions. Varying this threshold shifts the operating point: lower thresholds classify more samples as positive (\textit{e.g.}, video-streaming), increasing TPR/Recall but also FPR; higher thresholds classify fewer samples as positive, reducing TPR while typically lowering FPR. In the running example, an overly low threshold would label nearly all flows as video-streaming causing high FPR, whereas an overly high threshold would label nearly all flows as other traffic causing low TPR. Plotting TPR against FPR across thresholds creates the Receiver Operating Characteristic (ROC) curve, which is shown in \Cref{fig:roc}. The Area Under the Curve (AUC) metric is then calculated as its integral, providing a summary of performance. A higher AUC value indicates the superiority of a classifier and vice versa. 

\subsubsection{Multi-class scenarios}
The metrics described for binary classification are also applicable to multi-class tasks. Although the confusion matrix can become large, it naturally generalizes from $2\times2$ into an $N\times N$ form (where $N$ is the number of classes) that shows inter-class misclassifications and supports per-class evaluation.

Moreover, the multi-class Accuracy measure is defined as the ratio of correctly predicted observations to the total number of observations. Its formula can be simplified as:

\begin{equation*}
    Accuracy = \frac{\text{Correct Predictions}}{\text{All Predictions}}
\end{equation*}

Other metrics cannot be adapted to multi-class problems as straightforwardly, and some aggregation is necessary. There are three aggregation variants:

\begin{LaTeXdescription}
    \item[Macro:] Computes the metric independently for each class and then takes the unweighted average across all classes.
    \item[Weighted:] Computes the metric per class but weights each class by its number of true instances, giving more importance to classes with more samples.
    \item[Micro:] Aggregates the contributions of all classes by first summing the true positives, false positives, and false negatives across the entire dataset, and then calculates the metric globally.
\end{LaTeXdescription}

When reporting multi-class metric variants, it is necessary to specify the aggregation used by explicitly naming the metric (\textit{e.g.}, \textit{weighted precision} or \textit{macro $\text{F}_{1}$-score}). 

\subsection{Machine Learning Algorithm Selection}
Another challenge is deciding which ML model to train. There is no single \textit{best} algorithm for all traffic classification scenarios. The optimal choice requires balancing predictive performance, interpretability, computational cost, and ease of deployment. Characteristics of the dataset---such as class distribution and the number of target classes---also influence the selection. Historically, the common approach was to fit multiple models and choose the one with the highest accuracy, a process that was often time-consuming and required careful tuning, including feature selection for each algorithm. Today, with more complex models and longer training times, practitioners often select one state‑of‑the‑art method, such as boosting algorithms or neural networks, and focus on extensive optimization to achieve peak performance. In this section, we provide a concise overview of the algorithms employed. However, a detailed discussion of these ML methods lies beyond the scope of this paper. For comprehensive explanations and theoretical background, the reader is referred to~\cite{10.5555/1248547.1248548,Williams:2006:PPC:1163593.1163596,4738466}.

\subsubsection{Machine Learning Baselines}
While modern algorithms often achieve higher accuracy, classic ML models remain relevant. They serve as strong baselines for benchmarking the performance of more advanced techniques. Moreover, classic models are typically lightweight, fast to train and deploy, and offer transparent decision-making processes, which makes them particularly valuable in resource‑constrained environments or high‑stakes applications. We provide a brief introduction to three selected baseline methods. However, numerous other baseline approaches exist that are not discussed here, such as Logistic Regression and Naive Bayes.

\paragraph*{\textbf{Decision Tree}} It is a series of if/else conditions structured hierarchically in a tree format~\cite{10.1023/A:1022643204877}. During inference, a data sample is evaluated by traversing the tree from the root to a leaf node, systematically following the branches determined by the input features. The leaf node then assigns the sample to a specific target class. 

\paragraph*{\textbf{$k$-Nearest Neighbors ($k$-NN)}} This algorithm is an intuitive, instance-based model that classifies a new sample by finding the \textit{k} most similar samples in the training data (its \textit{nearest neighbors})~\cite{10.1109/TIT.1967.1053964}. The resulting label is assigned by the majority class in that neighborhood. 

\paragraph*{\textbf{Support Vector Machine (SVM)}} It operates by finding an optimal separating hyperplane between data points belonging to different classes in the feature space. The objective is to maximize the margin, which is the distance between this hyperplane and the nearest data points from each class, known as support vectors~\cite{10.5555/299094.299104}.

\subsubsection{Ensemble Methods}
The ensemble methods represent the current state-of-the-art and achieve high accuracy and robustness by combining predictions of many individual, weaker models (typically Decision Trees). The defining factor in the ensemble methods' properties is the learning procedure. There are three main learning procedures---bagging, boosting, and stacking.

\paragraph*{\textbf{Bagging}} The name is a short variant of bootstrap aggregating~\cite{breiman1996bagging}. It trains each model in an ensemble on different subsets of the training data and features. The final prediction is obtained by averaging outputs in regression tasks or applying majority voting for classification, which reduces variance and improves model stability. A typical example of Bagging is the Random Forest~\cite{10.1023/A:1022643204877}, where multiple Decision Trees are trained on bootstrap samples of the dataset, and each tree selects a random subset of features at each split. This design increases robustness to overfitting while preserving the ability to capture complex relationships between features and labels~\cite{10.1023/A:1010933404324}.

\paragraph*{\textbf{Boosting}} This is a family of ensemble learning algorithms in which models are trained sequentially, and each new model focuses on correcting the errors made by its predecessors. This iterative, error-correcting process typically results in high performance but is computationally demanding, as the sequential nature of training limits parallelization. Among the most widely used boosting implementations are XGBoost~\cite{chen2016xgboost} and LightGBM~\cite{ke2017lightgbm}, both known for their efficiency and scalability. In boosting ensembles, the final prediction is usually obtained by a weighted aggregation of the outputs from weak learners, most often Decision Trees. The exact method of combining these outputs varies across algorithms.

\begin{figure}[t]
	\centering
	\includegraphics[width=\columnwidth]{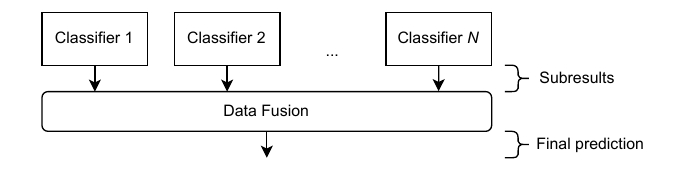}
	\caption{Stacked ensemble: outputs of multiple first-level classifiers are fused for the final prediction.}
	\label{fig:heterogeneous-method}
\end{figure}

\paragraph*{\textbf{Stacking}} This is a sequential ensemble method in which predictions from several base models (often called first-level learners) serve as input features for a higher-level model, known as the meta-learner (see~\Cref{fig:heterogeneous-method}). In the first stage, each base model is trained independently on the same dataset and predicts class probabilities or labels. These predictions are then used to construct a new feature set, on which the second-level model is trained to optimally combine the base first-level outputs. The process may be further extended across multiple stacked layers, with the top layer producing the final aggregated prediction. Each base learner can employ a different algorithm, which enhances the overall robustness and generalizability of the stacking ensemble~\cite{wolpert1992stacked}. 

\subsubsection{Neural Networks}
When the feature space is exceptionally complex and extensive training data are available, a neural network can be an appropriate choice. The simplest form of a neural network is the \textbf{Multilayer Perceptron (MLP)}, which remains the most commonly used architecture~\cite{almeida2020multilayer}. However, more sophisticated models, such as deep convolutional neural networks (CNNs), are also employed~\cite{luxemburk2023encrypted}. Despite the widespread enthusiasm surrounding neural networks in domains such as computer vision and natural language processing, their adoption in network traffic classification remains limited. In this domain, these models have notable limitations that often diminish their advantages.

Neural networks can learn rich patterns. However, they are less interpretable, demand considerable resources and robust training pipeline, while their design process is often long and error-prone. Their biggest advantage is the built-in feature engineering---the network itself identifies features in the raw data. Nevertheless, recent research also suggests that boosting algorithms can perform on par with complex convolutional neural networks, despite both being trained on raw packet sequence data without any tedious feature engineering process~\cite{luxemburk2023encrypted}. Therefore, traditional ML algorithms remain more commonly used in the field~\cite{alwhbi2024encrypted}.

One of the reasons for the relatively conservative use of neural networks in traffic analysis is their limited performance on tabular data~\cite{grinsztajn2022tree}, which is typical in this field. However, ongoing research continues to introduce specialized deep learning architectures for tabular data, such as TabNet~\cite{arik2021tabnet}, aimed at improving both performance and interpretability. Furthermore, since a substantial portion of data science research focuses on neural networks, we can expect their increasing prevalence in future network traffic studies, as these methods are likely to offer a wider range of advantages and enable better integration with other analytical systems. 

\subsubsection{Brief Summary on Algorithm Selection}

The choice of an algorithm always depends on the specific objectives of the classification task. \Cref{tab:model_comparison_revised} summarizes the key characteristics of the evaluated model families, showing their strengths and weaknesses. A comprehensive review~\cite{alwhbi2024encrypted} of the current state-of-the-art suggests that, for most practical scenarios, ensemble methods such as Random Forests and boosting algorithms (LightGBM or XGBoost) offer the most effective balance. They consistently combine high predictive accuracy, reasonable interpretability and manageable computational demands.

\begin{table*}[!ht]
\centering
\caption{A Comparative Guide to Supervised Learning Models for Feature-Based Traffic Classification}
\label{tab:model_comparison_revised}
\begin{tabular}{@{}p{3cm} p{4.5cm} p{4.5cm} p{4.5cm}@{}}
\toprule
\textbf{Model Family} & \textbf{Core Principle} & \textbf{Primary Strength} & \textbf{Primary Weakness / Trade-off} \\
\midrule
\textbf{Ensemble Methods} \newline (Boosting, Bagging, or Stacking) & Combines many weak learners (\textit{e.g.}, Classic Baselines) to create a single strong model via voting or sequential error-correction. & \textbf{State-of-the-art accuracy} on tabular data with excellent generalization. & Less directly interpretable than a single tree. Can be computationally intensive to tune. \\
\midrule
\textbf{Neural Networks} \newline (Multi-Layer Perceptron, Convolutional networks, TabNet) & Learns complex, non-linear feature combinations through multiple layers of \textit{neurons}. & Highest potential performance, especially on very large datasets (\textgreater 1M flows). & A \textit{black box} that is difficult to interpret. Requires significant data and careful, lengthy tuning. \\
\midrule
\textbf{Classic Baselines} \newline (Decision Tree, $k$-NN, SVMs) & Splits the feature space (Decision Tree), classifies based on similarity to neighbors ($k$-NN) or by finding a maximal separating hyperplane (SVM). & Simple, intuitive, and highly interpretable. Excellent for understanding the feature space. & Lower accuracy on complex problems. $k$-NN has high inference-time cost. \\
\bottomrule
\end{tabular}
\end{table*}

\subsubsection{Advanced Deep Learning Approaches}
\label{sec:ml-advanced}

In this section, we present several techniques that represent the current state-of-the-art in research. While many of these methods were initially developed and applied in domains outside of network traffic analysis, their effectiveness has also been demonstrated within networking contexts. For each technique, we provide a high-level overview explaining its core principles and potential benefits, along with selected references for further reading.

\paragraph*{\textbf{Meta-Learning}} This method is also known as learning to learn. It is a subset of ML where models are trained to adapt to new tasks autonomously. Unlike traditional supervised learning, where a model is trained for a single specific task, meta-learning involves training across multiple tasks, each with its own dataset. The goal is to develop a generalized model that can quickly adjust to new, previously unseen tasks using prior knowledge~\cite{finn2017model}.

An example of meta-learning is few-shot learning (FSL)~\cite{wang2020generalizing}. Traditional supervised learning usually requires hundreds or thousands of labeled examples to train a model. In contrast, few-shot learning aims to achieve high classification accuracy using only a few labeled training samples per task. The term $n$-shot learning refers to the number of labeled examples available per class, for example, one-shot learning (one example per class) or even zero-shot learning, where the model is expected to generalize without any labeled examples for the new task, often requiring specialized strategies.

Few-shot learning is particularly valuable in domains where labeled data are scarce or expensive to obtain~\cite{wang2020generalizing}. In network traffic classification, for instance, capturing and labeling real-world malware communication samples can be difficult and time-consuming. Meta-learning approaches like FSL help address this challenge by reducing the reliance on large labeled datasets while maintaining good performance.

\paragraph*{\textbf{Metric Learning}} This method, discussed in~\cite{kaya2019deep}, internally uses a distance metric that aims to put similar objects close together and subsequently increase the distance between dissimilar objects. The distance metric is learned during the training phase. Popular distance metrics include Euclidean (L2) and Mahalanobis distance. The Mahalanobis distance~\cite{davis2007information} is a Euclidean distance after a linear transformation of the feature space. A $k$-NN classifier with a custom distance metric can also be called a metric learning method.

Metric learning is often associated with deep learning, commonly referred to as deep metric learning. In this approach, deep neural networks are trained with specialized loss functions to transform data into a new embedding space. The objective remains the same as in classical metric learning: to embed similar samples close together while pushing dissimilar samples apart, typically based on class labels.

The key difference is that deep metric learning learns the data transformation function and maps raw input data into an embedding space where distance computations are meaningful. Classical metric learning usually operates in the original feature space, adjusting or learning the distance function without modifying the data representation itself.

\paragraph*{\textbf{Large Language Models (LLMs)}} These models are increasingly being explored for TC tasks, motivated by their demonstrated success in pattern recognition across diverse domains. A recent survey~\cite{bui2024systematicLLM} explored their applicability for intrusion detection and compared their performance to the traditional ML models. The study revealed that few-shot prompting, where a small number of labeled classification examples are provided within a prompt, does not achieve sufficient accuracy to replace current IDPS. Moreover, it also states that fine-tuned LLMs achieve high accuracy for known attacks, but their generalization capabilities are limited, as unknown attack detection experiences significant performance drops.

\paragraph*{\textbf{Transfer Learning and Domain Adaptation}}
A model trained on network data from one organization (the source domain) often exhibits reduced performance when applied to another (the target domain)~\cite{weiss2016surveyTransferLearning}. Transfer learning targets this issue by reusing knowledge from one domain and adapting it to another. When target domain labels are unavailable, unsupervised domain adaptation can be used to align feature distributions across domains.

Importantly, the traffic classification task in the source and target domains does not need to be identical. For example, a model may be pre-trained on a large, generic dataset using self-supervised metric learning to learn general traffic properties and then fine-tuned on a smaller dataset from the target domain. This approach often yields better performance than training solely on the limited target data.

\subsection{Model Training and Hyperparameter Tuning}

\subsubsection{Model Training}
Once a model family is selected, the training phase begins. This is not a single action but an iterative process of teaching the model and refining its configuration to achieve the best possible performance on unseen data. The core challenge is to find the sweet spot between two types of error: \textit{bias} and \textit{variance}. A model with high bias is too simple and fails to capture the underlying patterns in the data (underfitting). A model with high variance is too complex and learns the noise specific to the training set, failing to generalize to new data (overfitting). This trade-off is managed through the model's hyperparameters. It is essential to understand the difference between hyperparameters and parameters. 

The internal parameters are values learned automatically during training. The purpose of training is to adjust these parameters using an optimization algorithm to minimize a loss function---a measure of the model’s error---on the training data. In contrast, hyperparameters are externally defined configuration values that shape both the learning process and the model architecture. They determine the model's capacity, which influences how aggressively it can reduce the training loss. By tuning hyperparameters, we manage the bias–variance trade-off, balancing the model's ability to generalize against its tendency to overfit. For example:

\begin{itemize}
    \item For a \textbf{Random Forest}, `max depth` controls the complexity of each tree. A very large depth can lead to overfitting, while a very small depth can lead to underfitting.
    \item For a \textbf{Boosting algorithm}, the `learning rate` controls how strongly each new tree corrects the previous ones. A high rate can lead to overfitting, while a low rate may require more trees.
    \item For a \textbf{Multi-Layer Perceptron}, the `number of hidden layers` and `neurons` per layer directly control the model's capacity to learn complex patterns.
\end{itemize}

\subsubsection{Hyperparameter Tuning}

Since optimal hyperparameter values cannot be learned directly from the training data, they must be discovered through a robust tuning process using the validation set or k-fold cross-validation procedure (see~\Cref{subsec:data-partitioning}). 

The entire process is often highly experimental---setting hyperparameters, fitting the model, and observing the performance metrics. It can be viewed as an optimization problem; thus, there are many automated search strategies to explore the hyperparameter space effectively. Common approaches include Grid Search, which exhaustively tries every possible combination of hyperparameters provided, or Random Search. More advanced approaches include Bayesian Optimization, which samples combinations based on a probabilistic model to efficiently focus on the most promising regions of the hyperparameter space.

Hyperparameter tuning is an optimization task that may involve an extended search for the global optimum. Once the search is completed---or sufficient search rounds have been performed---the configuration yielding the highest average cross-validation score is selected. The model is then retrained on the entire training dataset using these optimal hyperparameters, producing the final version.

\subsection{Explainability and Interpretability}
\label{sec:explainability}

In recent years, the demand for explainable AI has risen exponentially~\cite{Weber2021}. Still, ML models are often referred to as \textit{black boxes} because their decision-making processes can be extremely difficult to explain, or in some cases, even impossible. However, it is essential to understand the reasoning behind certain predictions. This is especially important in TC, where ML is used for the detection of security threats and abnormalities in traffic. 

Explainability is often mistaken for interpretability, but the two refer to different concepts: explainability refers to the ability to understand \textit{why} a model output the prediction it did for a certain input. When a model has high explainability, we can understand what led the model to the prediction based on the input. On the contrary, interpretability refers to the inner functioning of models. It describes \textit{how} a model made the prediction, what inner mechanisms it uses, and how it works in general. However, researchers in the field do not always agree on these definitions, as the boundary between them is not always clear.

In this section, we summarize the most important methods for both interpretability and explainability, with a particular focus on traffic classification. 

\subsubsection{Taxonomy}

Methods can be divided into \textit{model-agnostic} (not dependent on the underlying model) and \textit{model-dependent} (specific for the model). Model-agnostic methods can be applied to any ML model, regardless of its architecture, and work with any set of inputs and outputs. In contrast, model-specific (or model-dependent) methods are tailored to a particular algorithm and can only be used with the models they were designed for.

We can further divide these methods into either \textit{local} or \textit{global}. Local methods aim to explain individual predictions by revealing why the model produced a specific output for a given input. For example, the Shapley values method provides the contribution of each feature to one particular sample classification. Global methods, on the other hand, provide insight into the model’s overall behavior, reasoning patterns, and global decision boundaries.

\subsubsection{Interpretability Methods}
\label{sec:int-approaches}
Two types of interpretability methods are described in this section: model-dependent methods that use inherent interpretability properties of ML models, and feature-based methods that use model inputs to increase visibility into its inner functioning.

\textbf{Model-dependent methods} use interpretable properties that are inherently built into the ML models themselves. Each method is applicable only to certain model families or algorithms. In this section, we summarize the most important methods:
\begin{enumerate}
    \item \textbf{Gini importance}, also known as Mean Decrease in Impurity (MDI), is commonly used with tree-based algorithms. The impurity in this context is a degree of class heterogeneity within a node---impurity of a node is zero when all samples in the node belong to the same class. Gini importance of a feature is computed as the total decrease in impurity it induces across all splits in the tree where it is used. In simple terms, features with higher Gini importance are generally more effective at partitioning the data. However, this metric can be biased: features with many unique values (high cardinality) or continuous features may receive disproportionately high importance scores, even if they do not genuinely enhance predictive performance.

    \item \textbf{Decision Trees} can be visualized. A tree-based graph with split conditions and annotated leaves explains the inner reasoning of the model. A path in the tree for a certain input can also be visualized, serving as both a global and local method. Moreover, it was previously used to craft rules (based on the exact path in the tree) for DDoS mitigation~\cite{zadnik2022}.

    \item \textbf{Linear Regression} uses a weighted sum of the input features. For classification, logistic regression can be interpreted by examining its coefficients, which characterize feature influence on the decision boundary.

    \item \textbf{Logistic Regression} has a linear decision boundary, but the relationship between input features and predicted probability is non-linear (sigmoidal). The coefficients represent the change in log-odds: increasing feature $x$ by one unit multiplies the odds by $\exp(w_x)$. Thus, examining the coefficients still provides interpretable insights into feature effects.

\end{enumerate}
\textbf{Feature-based methods} use input features to interpret model outputs. Such methods evaluate how each feature contributed to the prediction. The methods described in this section are model-agnostic as they can be used for any algorithm. The list below describes well-known feature-based methods:

\begin{enumerate}
    \item \textbf{Correlation} between each feature and the target variable can be calculated and used to select features with the highest correlation coefficient. More correlated features have a higher impact on the target variable.
    \item \textbf{Single-Variable Prediction} measures the importance of each feature by using only this feature to predict the target variable. Features with lower importance will produce poor predictions.
    \item \textbf{Feature importance} quantifies the influence of each feature on the predictions. Features with higher scores contribute more significantly to the decisions, while scores near zero suggest minimal or no contribution, indicating that such features could be removed without affecting performance. A complementary approach is permutation importance, which evaluates feature relevance by randomly shuffling the values of a specific feature. If the feature is important for classification, this shuffling will cause a noticeable drop in model performance.
\end{enumerate}

\subsubsection{Explainability Methods}
This section provides an overview of the most important explainability methods. The described methods provide justifications and reasoning behind model predictions. Many methods focus on explaining individual predictions; however, explanations can in some cases be aggregated to draw overall conclusions about the model.

\textbf{Partial Dependence Plot} (PDP)~\cite{friedman2001pdps} illustrates how the values of a specific feature influence the model's predictions. The \textit{x}-axis represents the selected feature's values, while the \textit{y}-axis shows the corresponding average prediction. To compute the PDP, a range of values for the chosen feature is selected. For each value, a modified dataset is created by setting that feature to the current value across all samples, keeping the other features unchanged. The model predictions for this modified dataset are then averaged to obtain the \textit{y}-axis value. The main limitation of PDP is its assumption of feature independence, which can lead to misleading results if the features are correlated. Additionally, PDP computation can be computationally expensive, especially for large datasets or complex models.

\begin{figure}[ht]
	\centering
	\includegraphics[width=1\columnwidth]{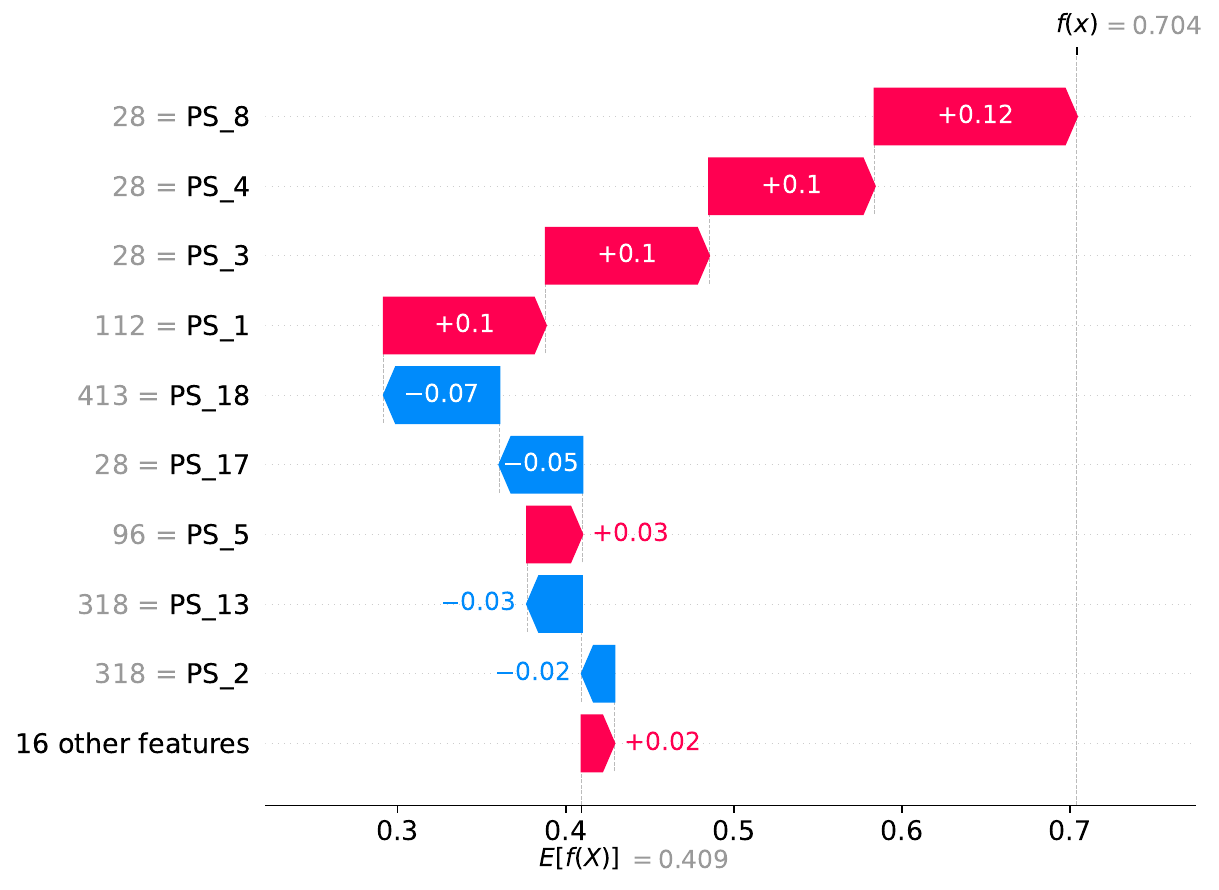}
    \caption{SHAP waterfall plot showing how each feature contributed to a specific prediction. Feature names and their observed input values are shown on the left; horizontal bar lengths represent SHAP values (feature contributions).}
	\label{fig:shap_waterfall}
\end{figure}

\textbf{SHAP (SHapley Additive exPlanations)}~\cite{lundberg2017unified} is a model-agnostic explainability method based on Shapley values from cooperative game theory~\cite{shapley1953value}. SHAP assigns each feature a contribution score for a specific prediction by treating features as players in a cooperative game. Each SHAP value represents a feature's average marginal contribution to the prediction across all possible subsets of features. In other words, it quantifies how much the observed value of a particular feature influences the prediction of class $y$ compared to the average baseline prediction for $y$, as~\Cref{fig:shap_waterfall} shows.

While SHAP is primarily used for local explanations of individual predictions, aggregating SHAP values across many samples enables global explanation. By computing the mean absolute SHAP value for each feature over the entire dataset, one can assess the overall importance of features in the model's decision-making process. SHAP satisfies several desirable theoretical properties such as local accuracy, consistency, and missingness~\cite{lundberg2017unified}. However, calculating exact Shapley values is computationally expensive, especially for models with many features, necessitating approximation methods such as Kernel SHAP or Tree SHAP for practical applications.

\textbf{Local Interpretable Model-agnostic Explanations (LIME)}~\cite{lime} uses an interpretable surrogate model (often linear) to locally approximate a black-box model. LIME perturbs a given input sample, queries the original model to obtain predictions, and then fits the surrogate model on the resulting synthetic dataset (weighted by proximity to the original sample). The explanation is derived from this local surrogate model, which is not intended to approximate the black-box model globally.

\textbf{Ablation Studies} are inspired by methodologies from neuroscience, where controlled damage to specific tissues is performed to observe its effects on the nervous system. This approach helps identify the role and importance of individual components within complex biological systems.

In ML, ablation studies serve a similar purpose by systematically removing or disabling certain components of a model to evaluate their contribution to overall performance~\cite{meyes2019ablation}. After removing or modifying a component of the model or training pipeline, the system is re-evaluated to observe changes in performance. This process provides insights into the specific role and contribution of the altered part to the overall system behavior. Ablation studies are regularly used for multiple purposes:
\begin{enumerate}
    \item Understanding model architecture: For example, removing an inner layer of a deep neural network can reveal its role in feature extraction or abstraction.

    \item Feature contribution analysis: Eliminating certain input features helps identify which features are crucial for the model's decision-making process.

    \item Assessing robustness: By disabling components or adding noise, researchers can test how resilient the model is to missing or corrupted information.

    \item Validating design choices: Ablation helps verify that added model complexity actually improves performance, rather than introducing redundant elements.
\end{enumerate}

\subsection{Machine Learning Model Deployment}
Deploying a model from an experimental setting into a real operational production system is a high-risk procedure. The model's properties and performance must be validated through experiments and thorough analysis, including explainability methods when applicable. Despite deployment being less studied than model development, in this section we summarize best practices for the final stage of the ML model lifecycle.

\subsubsection{Deployment Strategies}
A model can be deployed with several different strategies, each suitable for different operational requirements and risk tolerance. All of these methods are known from software engineering, but are applicable to ML pipelines as well.

\paragraph*{\textbf{Canary Deployment}} It gradually transitions the system from the old model to the new model. The model can be used only on a subset of traffic, reducing the risk of catastrophic failure. If early results are positive, the rollout increases incrementally until all traffic is analyzed by the new model.

\paragraph*{\textbf{Shadow Deployment}} It involves running a new ML model in parallel with the existing production system, but without impacting the production system itself. This approach is usually the first step in deployment, since it allows observation of the model under real conditions. After shadow deployment, the model can be directly switched to production, or gradual transition like Canary deployment can be used.

\paragraph*{\textbf{Blue-Green Deployment}} It maintains two identical production environments. At any given time, one environment serves active traffic while the other is prepared for updates. By switching between environments, teams can deploy new models and roll back instantly if issues arise.

\subsubsection{MLOps Lifecycle Frameworks}
The ML-based service will never be static. The ML model will evolve over time, and versioning systems and frameworks that link each model version to its corresponding dataset and preprocessing pipeline are essential in any production environment. Modern Machine Learning Operations (MLOps) frameworks such as MLflow or Kubeflow automate and monitor the entire lifecycle of ML models, from data preparation and training to validation and deployment. Each model can be traced back to the specific dataset and configuration used, ensuring reproducibility and transparency. Many frameworks also offer built-in deployment capabilities and provide standardized APIs for serving models in a scalable and maintainable way.

\subsubsection{Performance Observability and Drift Analysis}
Compared to other application domains, observing ML models' performance on network traffic analysis during the operational phase poses significant challenges. Obtaining ground-truth labels in real time is often infeasible. Without labeled data, it is challenging to detect concept drift or degradation in model accuracy over time. As a result, alternative evaluation strategies have to be used:

\paragraph*{\textbf{Input and Output Observation}} This is a technique for detecting data drift by continuously monitoring the distributions of input features and model outputs over time. When these distributions deviate significantly from their original baseline, it indicates a potential change in the underlying data patterns or target relationships. Such deviations often suggest that the model's assumptions about the data no longer hold, leading to degraded performance. In this case, retraining the model on the updated data becomes necessary to restore predictive accuracy.

\paragraph*{\textbf{Active Learning Principle}} This technique was already discussed in \Cref{subsub:DataCollection:PracticalStrategies} as a strategy for improving label quality. However, it is also an effective approach for obtaining the labels during the deployment phase. In this setting, the model actively identifies samples with low confidence and forwards them for manual or external labeling. The number of such samples can be configured to a relatively small amount, for example, only a few dozen per day. Such labeling can be performed by an auxiliary system that makes active verification requests or by an external, potentially costly and slow service capable of providing accurate labels. Despite the overhead, active learning enhances model observability and adaptability while keeping labeling costs manageable.

\subsubsection{Classification Reasoning}
\label{sec:exp-alerts-and-model-reasoning}
Many ML deployments in network traffic analysis are security-focused. When a model detects a potentially malicious flow, it generates an alert that is forwarded to a Security Information and Event Management (SIEM) system. Security analysts review the alert, correlate it with additional data, and decide on an appropriate response. Since these alerts can trigger critical disruptive actions (such as blocking traffic or isolating hosts), decision transparency is a key property to streamline the validation process.

Stacked ensembles (see~\Cref{fig:heterogeneous-method}) offer a practical approach to embedding decision reasoning within classifier outputs. When each first-level model is designed to detect a specific high-level property, its outputs can be aggregated and attached to the final alert. These intermediate results enrich the overall context, allowing analysts to interpret model decisions more effectively and respond faster to emerging threats. For instance, the system may issue an alert classifying the IP address as part of a Mirai botnet while providing supporting details such as detected command-and-control traffic, unusually small packets, suspicious TLS SNI domains, and signs of DDoS activity. This structured alert reasoning improves situational awareness and supports the analyst in the validation process.

\subsection{Challenges}
\label{sec:ml-challenges}
\subsubsection{Explainability and Alert Fatigue}
Explainability remains one of the greatest challenges in ML, not only within the TC domain but across all applications. While various methods exist to evaluate how individual features influence specific decisions or the overall model behavior, the internal workings of many ML models largely remain black boxes. An additional challenge arises when deploying these models in real-world settings. Building trust in the model and its outputs is critical. However, this trust is often lacking. As a result, even correctly predicted threats and alerts may be ignored by security operators, frequently due to alert fatigue.

Alert fatigue, recently surveyed in~\cite{tariq2025alert}, is a well-known issue where an overwhelming number of alerts desensitizes the personnel responsible for responding to them, often leading to missed or ignored alerts. This problem is particularly prevalent in healthcare, but it also affects security analysts working with ML-based detectors, especially in large networks. In the traffic classification domain, alert fatigue can result in overlooked correctly detected attacks and vulnerabilities, as analysts gradually lose trust in the deployed models.

To reduce the volume of traffic analyzed, security systems often employ traffic filters or limit monitoring to critical entities and sub-networks, areas where no alert should be overlooked. However, simply reducing the number of alerts is not enough; alerts must also be enriched with meaningful explanations to support effective decision-making. The current explanation methods lack deep knowledge about the underlying model and often expose sensitive information from its training data. This creates a significant challenge to practical deployment, especially with proprietary commercial models. One promising direction for the future is the use of Large Language Models, which can generate user-friendly, high-level explanations without revealing confidential model details.

\section{A Practical Guide with Jupyter Notebooks}
\label{sec:practical-part}

This tutorial includes a comprehensive guide to allow practical hands-on exercises apart from the theoretical knowledge presented in the previous sections. We prepared a series of Jupyter notebooks that implement the complete ML-based classification pipeline using real network traffic captures. The notebooks can be accessed at \cite{github}, and are designed to guide the reader through a rigorous experimental process, showcasing common pitfalls and complex realities that practitioners often encounter.

\subsection{Data Collection}
The first notebook demonstrates the principles and pitfalls of modern flow data collection, using the NFStream framework to teach a series of distinct concepts in traffic measurement. Its primary contribution is a blueprint for generating a comprehensive, ML-ready dataset that addresses real-world data integrity challenges.

\begin{description}[wide, font=\bfseries]
    \item[Key Concepts Taught:] The tutorial begins with foundational techniques, including bidirectional flow metering, statistical and sequential feature extraction, and integrated ntop Deep Packet Inspection (nDPI) based application labeling. 

    \item[Critical Challenges Addressed:] We provide guidelines on how to handle three critical challenges in Data Collection. First, we demonstrate the impact of \textit{NIC hardware offloading} (\textit{e.g.}, GRO), which creates spurious \textit{super packets} and poses a significant domain shift risk for ML models when deployed on NICs without hardware offloading enabled. Second, we solve the \textit{boundary flow problem}, showing how to correctly process split PCAP files to maintain statistical integrity and application context. Third, we introduce the \textit{`NFPlugin' system}, demonstrating its power for custom logic such as GeoIP enrichment and protocol-aware expiration.

    \item[Principal Outcome:] This notebook shows a complete data collection toolkit. The key lesson is that the capture environment and processing methodology may introduce artifacts that are not just noise but must be understood and controlled to ensure the validity of any subsequent ML model.
\end{description}

\subsection{Data Preparation}
The second notebook implements the data preparation pipeline, which is a key part of the design process. It provides a detailed method for converting large raw network traffic data into a clean, structured dataset suitable for ML.

\begin{description}[wide, font=\bfseries]
    \item[Key Concepts Taught:] The notebook demonstrates a systematic five-step process: exploration, quality assessment, dependency analysis, feature selection, and target analysis. It provides practical examples for identifying and quantifying a wide range of data quality issues.

    \item[Critical Challenges Addressed:] This module targets subtle but serious errors in network data preparation. First, it reveals that many missing application-layer values are structural rather than random, and that using these features directly can cause severe data leakage. Second, it demonstrates that data cleaning is inherently iterative, because initial filters can turn previously varying features into constants, which then requires revisiting earlier cleaning decisions.

   \item[Principal Outcome:] The notebook's most important lesson is the need for strict data preparation. It provides a data-based justification for aggressive quality filtering, where over \textit{62\% of the initial flows are programmatically removed} because they are identified as network noise (\textit{e.g.}, having fewer than 10 packets) or having low-confidence labels. 

\end{description}

\begin{table*}[ht]
    \centering
    \caption{Summary of the Practical Tutorial Notebooks and Their Contents}
    \label{tab:practical-summary}
    \begin{tabularx}{\textwidth}{@{} l X X c @{}}
        \toprule
        \textbf{Notebook}  & \textbf{Primary Objective} & \textbf{Key Techniques Taught} & \textbf{Paper Ref.} \\
        \midrule
        
        \href{https://github.com/FlowFrontiers/ml-flow-class-tutorial/blob/main/01-data-collection/01-data-collection.ipynb}{Data Collection} &
        To teach the principles and pitfalls of flow data collection using targeted examples. & 
        \texttt{NFStream} API, \texttt{nDPI} Labeling, SPLT Analysis, Plugin Extensibility (GeoIP, Slicing), Anonymization. & 
        \Cref{sec:data-collection} \\
        \addlinespace
        
        \href{https://github.com/FlowFrontiers/ml-flow-class-tutorial/blob/main/02-app-classification/02a-data-preparation.ipynb}{{Data Preparation}} &
        Systematically clean a large, real-world dataset and curate it for modeling. & 
        Iterative Filtering, Leakage Analysis, Target Variable Analysis, Strategic Dataset Curation (\texttt{Pandas}). & 
        \Cref{sec:data-preparation} \\
        \addlinespace
        
        \href{https://github.com/FlowFrontiers/ml-flow-class-tutorial/blob/main/02-app-classification/02b-comparative-modeling.ipynb}{{Comparative Modeling}} &
        Systematically compare ML algorithms and feature representations. & 
        Model Bake-Off (4 algos), \texttt{SMOTE} Trade-offs, Feature Importance, Many-Class Evaluation. & 
        \Cref{sec:machine-learning} \\
        \addlinespace
        
        \href{https://github.com/FlowFrontiers/ml-flow-class-tutorial/blob/main/02-app-classification/02c-advanced-optimization.ipynb}{{Advanced Optimization}} &
        Systematically validate, tune, and simplify a single model architecture (MLP). & 
        k-fold cross-validation, \texttt{GridSearchCV}, Feature Selection. & 
        \Cref{sec:machine-learning} \\
        \addlinespace
        
        \href{https://github.com/FlowFrontiers/ml-flow-class-tutorial/blob/main/03-explainability/03-explainability.ipynb}{{Explainability}} &
        To understand and trust a \textit{black box} model's decisions at a global and local level. & 
        Permutation Importance (w/ correlation handling), \texttt{SHAP}, \texttt{LIME}, PDPs. & 
        \Cref{sec:explainability} \\
        \bottomrule
        
    \end{tabularx}
\end{table*}

\subsection{Comparative Modeling}
This notebook contains the core of ML model design and exploration. Its primary contribution is a comprehensive, hands-on framework for model and feature selection in the context of encrypted traffic classification.

\begin{description}[wide, font=\bfseries]
    \item[Key Concepts Taught:] The notebook presents a complete ML workflow. It covers the training and evaluation of four distinct model families (\textit{Act 1}), the application of SMOTE to demonstrate the precision-recall trade-off, the use of feature importance for explainability, and the specific challenges of evaluating a many-class model.

    \item[Critical Challenges Addressed:] This module addresses two key questions in TC. First, it compares the predictive power of 55 engineered statistical features against a simple set of 25 raw packet sequence features (\textit{Act 2}). Second, it addresses the limitations of standard visualizations for multi-class problems by introducing programmatic evaluation techniques, such as plotting per-class $\text{F}_{1}$-scores and generating \textit{Top N Misclassification} reports (\textit{Act 3}).

    \item[Principal Outcome:] The experiments within this notebook yield two significant findings. First, tree-based ensembles (Random Forest, LightGBM) outperform linear models and simple MLPs for traffic classification. Second, Sequence of Packet Lengths and Times (SPLT) achieved higher accuracy than 55 statistical features, which shows the value of these raw data representations in the TC field.
\end{description}

\subsection{Advanced Optimization}
The fourth notebook shows the process of optimizing a candidate ML algorithm. It provides a complete workflow for model validation and hyperparameter tuning.

\begin{description}[wide, font=\bfseries]
    \item[Key Concepts Taught:] The notebook starts with establishing a baseline, then shows robust performance estimation with \textit{k-fold cross-validation} (and the correct use of \textit{pipelines} to prevent data leakage). Furthermore, automated hyperparameter tuning is done with GridSearchCV, and the notebook ends with a feature selection experiment to optimize the model further.

    \item[Critical Challenges Addressed:] This module tackles the trade-off between model complexity and performance. It focuses on optimizing model architectures to achieve high accuracy without incurring excessive computational cost or inference latency. Additionally, it addresses issues related to overfitting and generalization across different network environments.

    \item[Principal Outcome:] The notebook shows a practical demonstration of the impact of model simplification. While hyperparameter tuning provides some performance boost, the final feature selection experiment improves the score further, outperforming the complex model using all 55 features.
\end{description}

\subsection{Explainability}
The final notebook in the series addresses the critical \textit{black box} problem in ML, providing practical examples of model explainability. Its primary contribution is a hands-on demonstration of techniques for understanding and trusting a model's decisions. The notebook covers various ML model types, from inherently explainable models to more advanced ones.

\begin{description}[wide, font=\bfseries]
    \item[Key Concepts Taught:] The notebook covers both model-specific interpretability (\textit{e.g.}, visualizing a Decision Tree) and modern, model-agnostic explainability. It provides detailed examples of \textit{Permutation Feature Importance}, global explanations with \textit{Partial Dependence Plots (PDP)}, and local, per-prediction explanations using two state-of-the-art libraries: \textit{SHAP} and \textit{LIME}.

    \item[Critical Challenges Addressed:] This notebook focuses on the correct application of XAI (Explainable AI) techniques. Its most critical lesson is a practical demonstration of how \textit{multicollinearity (correlated features) can invalidate the explanations} of a naive permutation importance analysis. The notebook first shows the misleading results and then teaches the correct, advanced methodology.

    \item[Principal Outcome:] This notebook provides tools to explain model behavior. It teaches correct XAI application to avoid misleading conclusions about model logic.
\end{description}

\subsection{Summary of the Practical Framework}

Collectively, these five notebooks form an end-to-end workflow for network traffic classification. They guide the reader from the operational realities of packet capture and data integrity, through iterative data preparation, into systematic modeling experiments, culminating in advanced optimization and explainability techniques. \Cref{tab:practical-summary} provides a structured overview of these modules. This practical implementation serves as both a validation of the methodologies presented in the tutorial and as a reusable framework for researchers and practitioners.

\section{Conclusion}
\label{sec:conclusion}

This tutorial has presented a practical, end-to-end resource for developing and deploying ML-based network traffic classification systems. As modern networks increasingly rely on encrypted communications and diverse application behaviors, traditional port-based and deep packet inspection methods are often insufficient in many operational settings, especially under encryption and at scale. ML models operating on statistical flow features offer a practical and typically less intrusive approach to maintaining network visibility.

We systematically covered the complete pipeline required for building robust traffic classification systems across four core stages. In data collection (\Cref{sec:data-collection}), we explained how raw packet streams are transformed into structured flow records through metering systems, highlighting key engineering trade-offs between measurement fidelity and scalability. In data preparation (\Cref{sec:data-preparation}), we outlined a structured workflow for converting imperfect measurement data into high-quality, ML-ready datasets. In ML model development (\Cref{sec:machine-learning}), we introduced supervised classification principles and comprehensive performance metrics applicable to both binary and multi-class problems. Finally, in the practical component (\Cref{sec:practical-part}), we described the set of provided Jupyter notebooks designed for hands-on experimentation, enabling readers to bridge conceptual understanding with implementation practice. A recurring theme throughout the tutorial is practical rigor: obtaining reliable ground truth, recognizing measurement artifacts, and preventing data leakage when designing evaluation protocols.

This tutorial serves as a practical guide for practitioners building operational systems and as a reference point for researchers advancing the field. The process from packet capture to prediction and decision-making in production environments is complex, but understanding the full pipeline---with all its technical details and real-world constraints---is essential for developing robust and reliable systems. We encourage the community to build on this work, focus on the challenges identified throughout the tutorial, and continue improving both the scientific and practical aspects of network traffic classification.

\printbibliography


\begin{IEEEbiography}[{\includegraphics[width=1in,height=1.25in,clip,keepaspectratio]{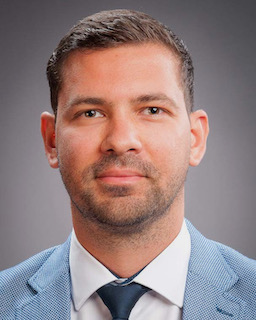}}]{Adrian Pekar} is currently a Senior Data Scientist at CUJO AI, where he develops ML-powered solutions for home networks, focusing on attack detection and encrypted traffic analytics. Previously, he held the position of Associate Professor 
at Budapest University of Technology and Economics, where he continues to teach part-time. 
His research interests encompass network traffic flow measurement, machine learning for traffic analytics, federated learning for traffic classification, and cybersecurity applications.
\end{IEEEbiography}

\begin{IEEEbiography}[{\includegraphics[width=1in,height=1.25in,clip,keepaspectratio]{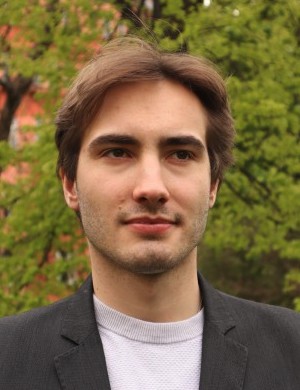}}]{Richard Plný} is currently pursuing a PhD degree at the Faculty of Information Technology, Czech Technical University in Prague. His research focuses on machine-learning-based network traffic classification and data fusion, with emphasis on explainability and deployability. He is also a researcher at CESNET, where he works on threat detection in large ISP-level networks, including the identification of malicious cryptomining activities.
\end{IEEEbiography}

\begin{IEEEbiography}[{\includegraphics[width=1in,height=1.25in,clip,keepaspectratio]{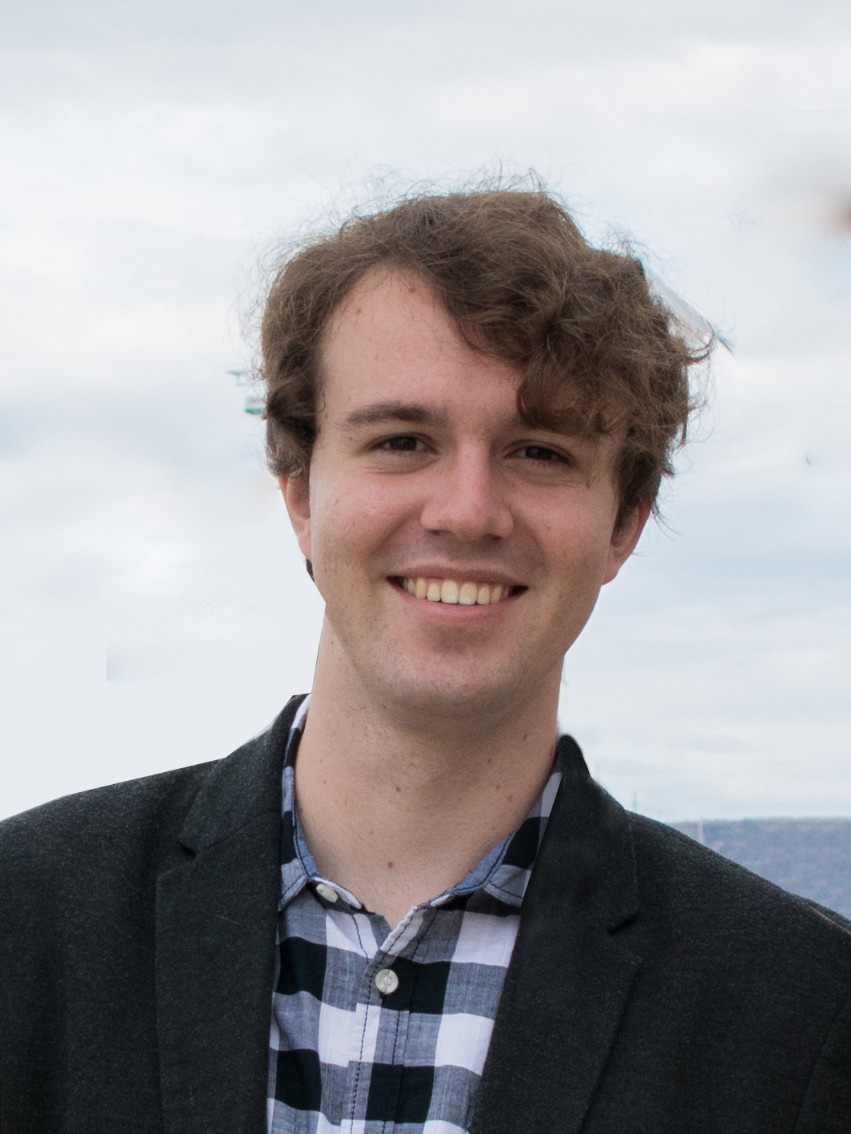}}]{Karel Hynek} is currently researcher and educator at the Faculty of Information Technology, Czech Technical University in Prague, where he is a core member of the Network Monitoring Laboratory (NETMON). His expertise lies in network security, with an emphasis on high-speed monitoring and ISP-scale protection systems. His research outputs, such as traffic classifiers, detectors, and data exporters, are actively deployed to support and protect a production ISP monitoring infrastructure.
\end{IEEEbiography}






\end{document}